\documentclass[letterpaper,12pt]{article}

\usepackage[ruled,vlined]{algorithm2e}
\usepackage{tikz}
\usetikzlibrary{decorations.pathreplacing}
\usepackage{comment}
\usepackage{tabularx} % extra features for tabular environment
\usepackage{float}
\usepackage{amsmath}  % improve math presentation
\usepackage[margin=1in,letterpaper]{geometry} % decreases margins
\usepackage{cite} % takes care of citations
\usepackage[final]{hyperref} % adds hyper links inside the generated pdf file
\usepackage{xargs}                      % Use more than one optional parameter in a new commands
\usepackage[utf8]{inputenc}
\usepackage[english]{babel}
\usepackage{graphicx}
\usepackage{subfigure}
\usepackage{tikz} 
\usetikzlibrary{arrows,fit,positioning}
\usetikzlibrary{matrix}
\usetikzlibrary{calc}
\usetikzlibrary{arrows}
\tikzstyle{int}=[draw, fill=white!8, minimum size=2em]
\tikzstyle{init} = [pin edge={to-,thin,black}]

\usepackage{geometry}
\definecolor{myLightGray}{RGB}{200,200,200}
\definecolor{myGray}{RGB}{160,160,160}
\definecolor{myDarkGray}{RGB}{144,144,144}
\definecolor{myDarkRed}{RGB}{167,114,115}
\definecolor{myRed}{RGB}{255,58,70}
\definecolor{myGreen}{RGB}{0,255,71}

\usepackage[normalem]{ulem}

\DeclareUnicodeCharacter{2212}{-}
\begin{document}

\title{Bayesian sequential data assimilation for COVID--19 forecasting}

\author{Maria L. Daza--Torres\footnotemark[1]\ \footnotemark[4]
\and
Marcos A. Capistr\'an\footnotemark[2]
\and 
Antonio Capella\footnotemark[3]
\and 
J. Andr\'es Christen\footnotemark[2]
}

\renewcommand{\thefootnote}{\fnsymbol{footnote}}
\footnotetext[1]{Department of Public Health Sciences, University of California, Davis, United States.\textit{mdazatorres@cimat.mx}}
\footnotetext[2]{Centro de Investigaci\'on en Matem\'aticas (CIMAT), Jalisco S/N, Valenciana, Guanajuato, 36023, M\'exico.\textit{marcos, jac at cimat.mx}}
\footnotetext[3]{Instituto de Matem\'aticas, UNAM, Circuito Exterior, CU, CDMX, M\'exico \textit{capella@im.unam.mx}}

\footnotetext[4]{Corresponding author}

\date{\today}
\maketitle

\begin{abstract}
We introduce a Bayesian sequential data assimilation method for COVID--19 forecasting. It is assumed that suitable transmission, epidemic and observation models are available and previously validated and the transmission and epidemic models are coded into a dynamical system. The observation model depends on the dynamical system state variables and parameters, and is cast as a likelihood function. We elicit prior distributions of the effective population size, the dynamical system initial conditions and infectious contact rate, and use Markov Chain Monte Carlo sampling to make inference and prediction of quantities of interest (QoI) at the onset of the epidemic outbreak. The forecast is sequentially updated over a sliding window of epidemic records as new data becomes available. Prior distributions for the state variables at the new forecasting time are assembled using the dynamical system, calibrated for the previous forecast. Moreover, changes in the contact rate and effective population size are naturally introduced through auto--regressive models on the corresponding parameters. We show our forecasting method's performance using a SEIR type model and COVID--19 data from several Mexican localities.
\end{abstract}

\section{Introduction}
\label{sec:intro}

\bigskip

In this paper we introduce a Bayesian sequential data assimilation method for COVID--19 forecasting.
Reliable model--based COVID--19 forecasting should be helpful to assist decision--making and planning for healthcare authorities. Consequently, the contributions of this work are aimed at building confidence in our methodology.

Compartmental epidemic models have proven to be adequate to assimilate epidemic data and making forecasts~\cite{asher2018forecasting,bertozzi2020challenges}. However, epidemic outbreak predictability is limited due to the influence of human behavior, incomplete knowledge of the virus's evolution, and weather~\cite{castro2020turning,wilke2020predicting}, as well as delay and under--reporting of new cases and deaths~\cite{krantz2020level,lau2021evaluating}. A practical compromise is to make probabilistic epidemic forecasts a few weeks ahead of time~\cite{brooks2020comparing,engbert2021sequential} using a model that accounts explicitly for data delay and under--reporting.

In this paper we assume that transmission, epidemic and observation models are properly postulated, previously validated and available. The transmission and epidemic models are coded into a dynamical system. The observation model depends on the dynamical system state variables and parameters, and is cast as a likelihood function. In Section~\ref{sec:seird} we use a SEIR type epidemic model with Erlang ~\cite{champredon2018equivalence} residence times in the exposed and infected compartments. We elicit prior distributions of the effective susceptible population size, dynamical system initial conditions, and infectious contact rates and use Markov Chain Monte Carlo to make inference and prediction of quantities of interest (QoI), such as hospital occupancy, at the onsetof the epidemic outbreak in a metropolitan area. Namely, at the time when community transmission starts in the metropolitan area being analyzed. As new data becomes available, we update the forecast sequentially over a sliding window of epidemic records. We assemble prior distributions for the state variables at the new forecasting time using the dynamical system calibrated for the previous forecast. Moreover, we introduce changes in the contact rate and effective population size naturally through auto--regressive models on the corresponding parameters. We argue that this is a natural approach to data assimilation with an epidemic model.

We show our forecasting method's performance using a SEIR type model and COVID--19 data from several Mexican localities. 

\subsection{Related work} 

Real time epidemic forecasting is an emerging research field~\cite{desai2019real}. Many forecast modeling efforts study how to address data under--reporting and delays~\cite{gibson2020real,engbert2021sequential}. Other efforts are directed at exploring what sources of information can be incorporated as covariates to make better forecasts. Mcgough {\it et al.}~\cite{mcgough2017forecasting} incorporate traditional surveillance with social media data to forecast Zika in Latin America. The RAPIDD ebola forecasting challenge~\cite{viboud2018rapidd} explored how to integrate different sources of data for Ebola forecasting. Hii {\it et al.}~\cite{hii2012forecast} use temperature and rainfall to forecast dengue incidence.

In a related work, \cite{capistran2021PLoSONE} present a COVID19 prediction model.  Using a SEIR type dynamical model, and including hospital dynamics and Erlang compartments~\cite{champredon2018equivalence} to properly model residence times, \cite{capistran2021PLoSONE} model and predict the COVID19 epidemic in the Mexican 32 states and several metropolitan areas, from the epidemic onset in Mexico in March 2020 (and until February 2021, see \url{https://coronavirus.conacyt.mx/proyectos/ama.html} (in Spanish), model \textit{ama2}).  However, fitting the whole of the epidemic, to infer initial state values, for an epidemic lasting several months, ceases to be useful and adds to the numerical complexity and reliability of the system.  In fact, given the generation interval of COVID19, data beyond one month in the past should start to have less importance for current nowcasting and predictions.

\subsection{Contributions and limitations}

The probabilistic forecasting method introduced in this paper allows us to forecast the incidence of new cases and deaths one to four weeks in advance 

Once we are near or after a local incidence maximum, our forecasting method disentangles the role of infectious contact rate and effective population size. Other quantities of interest such as hospital occupancy can be calculated as a byproduct of the forecast using suitable renewal equations. More general data analysis, e.g. by age groups, is not presented in this work. However our results may be applicable on those cases, provided suitable transmission and epidemic models are available.

This manuscript is organized as follows. In Section~\ref{sec:methods} we make a summary of the modeling decisions taken to implement our forecasting method. In Section~\ref{sec:seird} we apply our method to COVID--19 epidemic data. Finally, in Section~\ref{sec:results} we present the analysis of the Mexico City data. Other examples are provided in the supplementary material.

\section{Bayesian Sequential Forecasting Method}
\label{sec:methods}

\begin{figure}[h!]
\vspace*{5mm}
\centering
\tikz \node [scale=1.4, inner sep=0] {
\begin{tikzpicture}[%
    every node/.style={
        font=\scriptsize,
        text height=1ex,
        text depth=.25ex,
    },
]
\draw[->] (0,0) -- (11,0);
\foreach \x in {1,2.8,6,8.5}{
    \draw (\x cm,3pt) -- (\x cm,0pt);
}

% place axis labels
\node[anchor=north] at (1.2,-0.1) {$t_k$};
\node[anchor=north] at (3.25,-0.1) {$t_{k+1}=t_{k}+n$};
\node[anchor=north] at (6.7,-0.1) {$t_k+L+D$};
\node[anchor=north] at (9.4,-0.1) {$t_k+L+D+F$};

\fill[myDarkGray] (1,0.6) rectangle (4,0.8) node[anchor=south,midway,above=6pt, black] {training period};
\fill[myRed] (4,0.6) rectangle (6.0,0.8) node[anchor=south,midway,above=6pt, black] {delay period};
%\fill[myRed] (9,0.25) rectangle (11,0.4);
\fill[myGreen] (6.0,0.6) rectangle (8.5,0.8) node[anchor=south,midway,above=6pt, black] {forecasting period};

% draw scale below
\fill[myLightGray] (1,-0.6) rectangle (2.8,-0.8);
\fill[myDarkGray] (2.8,-0.6) rectangle (4,-0.8);
\fill[myDarkRed] (4,-0.6) rectangle (5.8,-0.8) node[anchor=south,near start,below=8pt, black] {new training period};
\fill[myRed] (5.8,-0.6) rectangle (7.8,-0.8);
\fill[myGreen] (7.8,-0.6) rectangle (10.3,-0.8);
\draw[dashed,black] (2.79,-0.85) -- (2.79,0.85) -- (5.8,0.85)-- (5.8,-0.85)--(2.8,-0.85);
\end{tikzpicture}};
\caption{{\bf Bayesian Sequential data assimilation.} We propose a Bayesian filtering method that predicts along the dynamical system~\eqref{eq:model} evaluated in sample points of the posterior distribution $\pi_{P^{(k)}|Z^{(k)}}(p^{(k)}|z^{(k)})$ in the current forecasting window $[t_{k},t_{k}+L+D+F]$.\label{fig:forecasting_timeline}}
\end{figure}
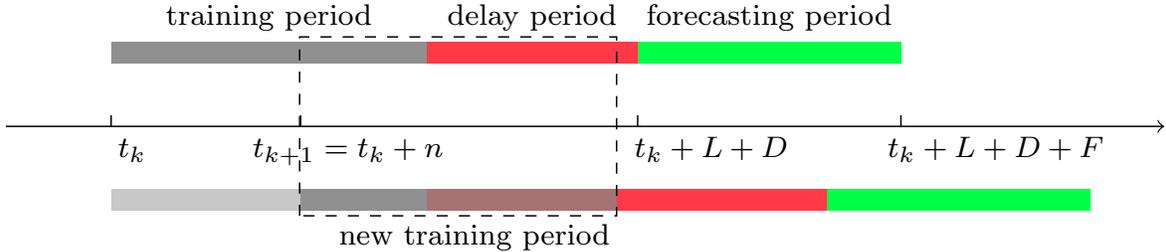

Let us assume that community transmission starts at time $t=t_0$ at the metropolitan area where the outbreak is being analyzed.
Set $k=0$ and denote by $[t_k,t_k+L]$ the learning period. Namely, the period when we collect epidemic records $z^{(k)}$ to create a forecast. In the example presented in Section~\ref{sec:seird}, these epidemic records are new hospital admittances and deaths.
The delay period is $[t_k+L,t_k+L+D]$, i.e. the period when epidemic records are not mature and may include delays in reporting.
The forecasting day is $t_k+L+D$. We refer to $[t_k+L+D,t_k+L+D+F]$ as the forecasting period, and $[t_k,t_k+L+D+F]$ is the forecasting window as illustrated in Figure~\ref{fig:forecasting_timeline}. 

Let $x(t)=(S(t),E(t),I(t),...)^T$ denote the time--dependent vector of state variables. We shall assume that the epidemic and transmission models are posed as an initial value problem for a nonlinear system of ordinary differential equations

\begin{equation}
    \label{eq:model}
    \begin{split}
    \dot{x}(t)&=f(x(t),\theta_{k})\\
    x(t_k)&=x_{k},
    \end{split}
\end{equation}
where $t_k$ and $x_k$ denote respectively the initial time and state in the forecasting window $[t_k,t_k+L+D+F]$, and $\theta_{k}$ is a vector of model parameters (e.g. contact rate $\beta$, effective population size $\omega$, etc.) used to calibrate model~\eqref{eq:model}. We shall denote $p^{(k)}=(x_k,\theta_{k})$ the joint vector of initial conditions and model parameters to be inferred. 

If $k=0$, we postulate a prior distribution $\pi_{P^{(k)}}(p^{(k)})$, a likelihood $\pi_{Z^{(k)}|P^{(k)}}(z^{(k)}|p^{(k)})$ and use equation~\eqref{eq:model} and samples obtained through Markov Chain Monte Carlo of the corresponding posterior distribution $\pi_{P|Z}(p^{(k)}|z^{(k)})$ to make a probabilistic prediction of $x(t)$ in the forecasting period $t\in[t_k+L+D,t_k+L+D+F]$. Afterwards, we update the forecasting window by setting $t_{k+1}=t_{k}+n$, where $n$ is the number of days until the next forecast. We assemble a new prior distribution $\pi_{P^{(k+1)}}(p^{(k+1)})$ for the model parameters $p^{k+1}$ in the new forecasting window $[t_{k+1},t_{k+1}+L+D+F]$ using the predicted values of $x(t)$ at $t=t_{k+1}$ obtained with equation~\eqref{eq:model} and samples of the posterior distribution $\pi_{P^{(k)}|Z^{(k)}}(p^{(k)}|z^{(k)})$ of the previous forecast. Model parameters $\theta_{k+1}$ have an autoregressive prior distribution in terms of $\theta_{k}$. Finally, we set $k\leftarrow k+1$ and repeat the above process to create a new forecast.

\begin{algorithm}[ht]
\SetAlgoLined

\begin{enumerate}
\item[Input.]  Initial time ($t_0$), data ($z^{(k)}$) for $k=0,1,...$, learning period size size ($L$), number of days to forecast ($F$), number of days to move the forecasting window $n$, and number of the delays days ($D$), 

\item[Output.]
\begin{itemize}
\item Posterior distribution $\pi_{P^{(k)}|Z^{(k)}}(p^{(k)}|z^{(k)})$ for $k=0,1,...$
\item Prediction of QoI, e.g. hospital occupancy, report of new cases, etc, in the forecasting period $[t_k+L+D,t_k+L+D+F]$ for $k=0,1,...$
\end{itemize}

\item [Step 1.] If $k=0$:\\ 
\hspace*{8mm} Set a prior distribution for  $p^{(k)}$, $\pi_{P^{(k)}}(p^{(k)})$\\
\hspace*{8mm} Go to Step 3\\
else: \\
\hspace*{8mm}Go to Step 2

\item [Step 2.] For $k>0$, the prior distribution  for $p^{(k)}=(x_k,\theta_k)$ is set using the MCMC output of the period $k-1$:
\begin{itemize}
\item For the $k-$initial state ($x_k$), the MCMC output of the state variable at time $t_0 + nk$ is fitted a known distribution. 
\item For the model parameters $\theta_k$, the MCMC output of $\theta_{k-1}$ is fitted a known distribution.
\end{itemize}

\item[Step 3.] Compute posterior distribution, $\pi_{Z^{(k)}|P^{(k)}}(z^{(k)}|p^{(k)})$
\item[Step 4.] Forecast QoI up to time $t = t_0 + nk + L + D + F $\;
\item[Step 5.]  Save the MCMC output for the next forecasting time.
\end{enumerate}
\caption{Bayesian sequential data assimilation for COVID-19 forecasting \label{Alg:forecasting}}
\end{algorithm}

The Bayesian sequential data assimilation method consists of three parts; a dynamical system that codes the transmission and an epidemiological model, a probabilistic model for the observed incident cases and deaths, and an informed prior distribution for the parameter space in each forecasting period. In Section~\ref{sec:seird}, we show how to postulate each model component for a forecasting model of covid-19 using data from several Mexico localities.

\section{Example: A SEIR type model}
\label{sec:seird}
\subsection{Dynamical model}
We consider a variation on the SEIRD epidemic model for susceptible, exposed, infectious, removed, and dead individuals. We have added a compartment for unobserved infectious individuals.

We assume that the total population of the metropolitan area being analyzed is $N$. We assume further that there is only a small number of infected individuals at the onset of community transmission. Susceptible individuals $S$ become exposed $E$ with force of infection $\lambda$.
The transmission model is coded into $\lambda$ as follows, we assume that only unobserved ($U$) and observed ($O$) infectious individuals spread the infection, that is
$$\lambda=\frac{(U + k O)\beta}{N},$$
where $\beta$ is the infectious contact rate. We have assumed that the contact rate for observed infectious is a factor ($k$) of the contact rate for unobserved infectious. A fraction $f$ of exposed individuals proceeds to the observed infected class ($O$) at rate $\sigma_1$, while the remainder goes directly to an unobserved infective stage ($U$), also at rate $\sigma_1$. Individuals leave the infectious class at rate $\sigma_2$, with a fraction $1-g$ recovering and going to the removed class ($R$) and the remainder ($g$) dying of infection. Unobserved go the removed stage at rate $\gamma$. We split the $E$, $I$, and $O$ compartments into two sub-compartments to model residence rates explicitly as Erlang distributions \cite{champredon2018equivalence}, see Table \ref{tab:Erlang}.

The dynamics of the epidemic process is governed by the following nonlinear system of ordinary differential equations

% Kull2020 para ver mas del SEIRD
\begin{eqnarray*}\label{eq:SEIARD}
\dot{S}&=& -\lambda  S \\
\dot{E}&=& \lambda S -\sigma_1 E \\
\dot{O} &=& f\sigma_1 E - \sigma_2 O\\
\dot{U} &=& (1-f)\sigma_1 E - \gamma U\\
\dot{R} &=& (1-g)\sigma_2 O + \gamma U\\
\dot{D} &=& g\sigma_2 O,
\end{eqnarray*}
with initial conditions;
$E(0)= E_0,\; O(0)=O_0,\; U(0)=U_0,\; R_0=R(0), \; D_0=D(0)$, and $S(0)=N - E_0 - O_0 - U_0 - R_0 - D_0$. Here $N = S + E + O +U + R + D$. 

\begin{figure}[h!]
\centering
\begin{tikzpicture}[node distance=2.cm,auto,>=latex']
   \node [int] (S) {$S$};
   \node [int] (E) [right =2cm of S] {$E$};   
    \node [int] (U) [above right=2cm of E] {$U$};            
    \node [int] (O) [below right=2cm of E] {$O$};                   
    \node [int] (R) [right=4cm of E] {$R$};                       
    \node [int] (D) [right=1.8cm of O] {$D$};                       

   {\small
    \path[->] (S) edge node {$\lambda$} (E);
    \path[->] (E) edge node  {$(1-f)\sigma_1$} (U);
    \path[->] (E) edge node  [below left]{$f\sigma_1$} (O);  
    \path[->] (U) edge node  {$\gamma$} (R);        
    %\path[->] (I) edge node  [above left]{ } (C);            
    \path[->] (O) edge node [below] {$g\sigma_2$ } (D);                    
    \path[->] (O) edge node [left]{$(1-g)\sigma_2$ } (R);  
    }                    
\end{tikzpicture}
\caption{A SEIR type model that into account both observed and unobserved infections.\label{fig:diag_SEIARD}}
\end{figure}
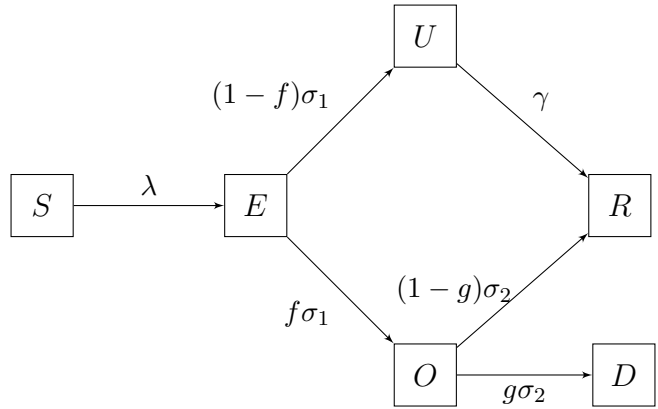

A flow diagram for the model is shown in Figure \ref{fig:diag_SEIARD}. 

\begin{table}
\caption{Average times and Erlang shape parameters for sub-compartments. \label{tab:Erlang}}
\begin{center}
\begin{tabular}{ c c c c c}
Variable & Rates & Average time & Erlang shape & Reference \\
\hline
$S$ & $\beta$ & Inferred & 1 & \\ 
$E$ & $1/ \sigma_1$ & 5 days & 2 &\cite{lauer2020,jiang2020does} \\  
$A$ & $1/ \gamma$  & 7 days & 2& \cite{long2020clinical}  \\ 
$I$ & $1/ \sigma_2$  & 14 days & 2&\cite{verity2020estimates,bi2020epidemiology} \\
\end{tabular}
\end{center}
\end{table}

\subsection{Model parameters}
The model has two kinds of parameters that have to be calibrated or inferred; the ones related to COVID-19 disease (such as residence times and proportions of individuals that split at each bifurcation of the model) and those associated with the public response to mitigation measures such as the contact rate Beta and the proportion of effective population size during the outbreak ($\omega$). Some of these parameters can be found in recent literature (see Table \ref{tab:Erlang}) or inferred from reported cases and deaths, but some remain mostly unknown. In the latter category, we have the fraction $1 − f$ of unobserved infections. We assume $1-f=0.2$, which means that $80\%$ of cases of symptomatic/asymptomatic infectious go unreported. 

\subsection{Observational model and data}
\label{sec:observational_model}
The observed data used to fit the model is based on time series of incident confirmed cases and deaths.
We consider daily deaths counts $d_i$ and its theoretical expectation that is estimated in terms of the dynamical model as
$$
\mu_D(t_i) = D(t_i) - D(t_{i-1}).
$$
Analogously, we consider daily case $c_i$ and its corresponding $\mu_c(t_i)$ given by the daily flux entering the $O$ compartment \cite{capistran2021PLoSONE}, namely

$$\mu_c(t_i) =\int_{t_{i-1}}^{t_i} f \sigma_1 E_2(t)dt$$ 
where $E_2(t)$ is the last state variable in the $E$ Erlang series. We calculate the above integral using a simple trapezoidal rule with 10 points.

\subsection{Estimating model parameters with MCMC}

We consider daily confirmed cases $c_i$ of patients with a positive test ($O$)  and daily report deaths $d_i$, for the area being analyzed. To account for over dispersed counts, we use a negative binomial (NB) distribution $NB(\mu, \omega,\theta)$ with mean $\mu$ and over dispersion parameters $\theta$ and $\omega$. For data $y_i$, we let
$$y_i\sim NB(p\mu(t_i),\omega,\theta),$$
with fixed values for the over dispersion parameters $\omega,\theta$ and an additional reporting probability $p$. We assume conditional independence in the data and therefore from the NB model we obtain a likelihood. \\

The parameters to inferred are the contact rate ($\beta$), the proportion of the effective population ($\omega$), the fraction of infected dying ($g$), and crucially we also infer the initial conditions for  $E(0)$, $O(0)$, $U(0)$, $R(0)$, $D(0)$. Letting $S(0) = \omega \cdot N - (E(0) +  O(0) + U(0) + R(0))$. We have all initial conditions defined and the model can be solved numerically to obtain  $\mu_D$ and $\mu_c$ to evaluate our likelihood.\\

%\subsection{Eliciting prior distributions}

%In the first forecast, we use $Gamma$ distributions for initial conditions $E_0$, $I_0$, $A_0$ with scale $1$ and shape parameter $10$ to model the initial conditions modeling the low, near to $10$, and close to zero counts for the number of infectious initial conditions. For the initial conditions $R_0$, $D_0$, we also use $Gamma$ distributions with scale and shape parameters equal to $1$. This because at the beginning of the outbreak, both parameters are close to zero. The prior distributions for the remaining parameters is summarized in  Table \ref{tab:PriorParams}.

Regarding the elicitation of the parameters' prior distribution for the first forecast, we use Gamma distributions for the initial conditions $E_0$, $O_0$, and $U_0$, with scale $1$ and shape parameter $10$. This for modeling the low, near to $10$, and close to zero counts for the number of initial infectious conditions. For the initial conditions $R_0$ and $D_0$, we also use Gamma distributions with scale and shape parameters equal to $1$. This because at the beginning of the outbreak, both parameters are close to zero. The prior distributions for the remaining parameters are summarized in  Table \ref{tab:PriorParams}.

To sample from the posterior, we resort to MCMC using the t-walk generic sampler \cite{christen2010twalk}. The MCMC runs semi-automatic, with consistent performances in most data sets.

\begin{table}[H]
\caption{Parameters and prior distributions used for Bayesian inference.\label{tab:PriorParams}}
\begin{center}
\begin{tabular}{ c c c }
Parameter & Prior distribution\\
\hline
Contact rate ($\beta$) & $LogNorm(1,1)$ \\
Fraction of infected dying ($g$) &  $Beta(1 + 1 / 6, 1 + 1 / 3)$\\
Proportion of the effective population ($\omega$)& $Beta(1 + 1 / 6, 1 + 1 / 3)$\\
\end{tabular}
\end{center}
\end{table}

\section{Results}
\label{sec:results}
This section displays the results to apply Algorithm \ref{Alg:forecasting} with covid-19 data set from Mexico's city metropolitan area. Further examples of other Mexican metropolitan areas can be found in the supplementary material. The data set reports the incident number of confirmed cases and deaths for each location at a daily frequency starting in early 2020.

\begin{figure}[ht]
\begin{center}
\subfigure[]{
 \includegraphics[scale=0.38]{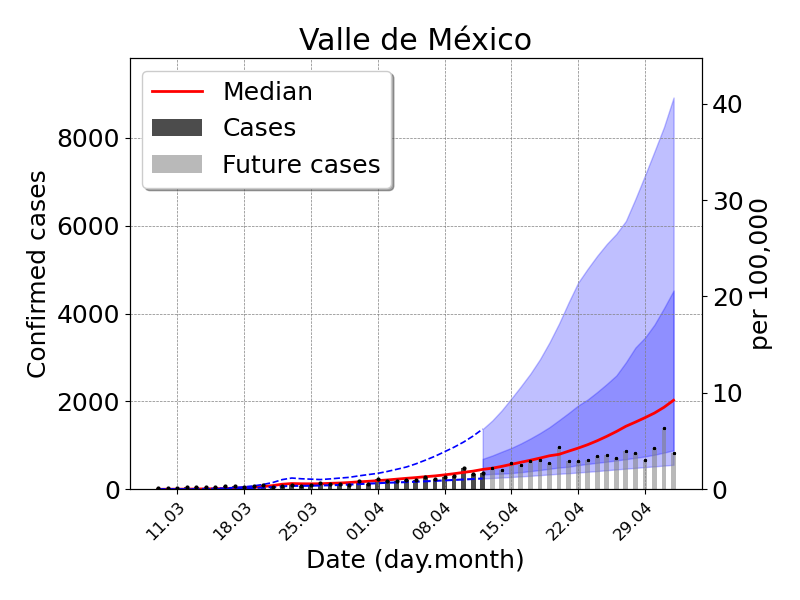}}
 \subfigure[]{ \includegraphics[scale=0.38]{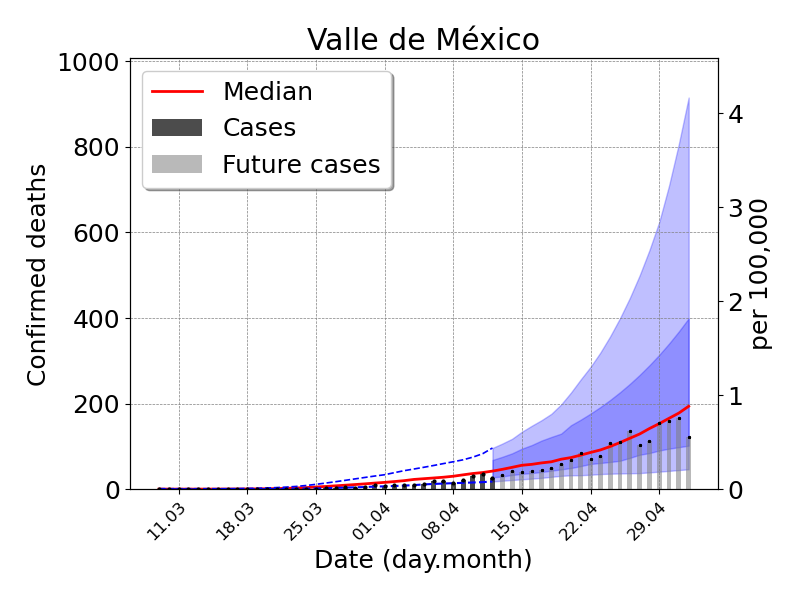}}
 \end{center}
\caption{Outbreak analysis for Mexico city metropolitan area, using data from March 8 to April 12. (a) Confirmed cases (b) Confirmed deaths.  Central red lines indicate the median incidence forecast. The darker shaded region indicates the interquartile forecast range, and the lighter shaded region indicates the 5–95th percentile range. All displayed forecast durations are ten days from the point of prediction.  Total population 21,942, 666 inhabitants.\label{fig:forecast1_VM}}
\end{figure}

The uncertainty in the early forecast is high because we do not yet know the effective size of the population participating in the epidemic. Further, these forecasts are prone to further errors given the uncertainty in disease spread parameters and the initial state of the disease. We use the Bayesian Sequential Forecasting Method to predict trajectories, given weekly updates. The model starts with inaccurately predicted trajectories, where the median of the trajectories overestimate the future data (See Figure \ref{fig:forecast1_VM}). Further, the prediction has a high initial cone of uncertainty. 

\begin{figure}[H]
\subfigure[Forecast 7, using data from April 27 to May 31.]{
 \includegraphics[scale=0.35]{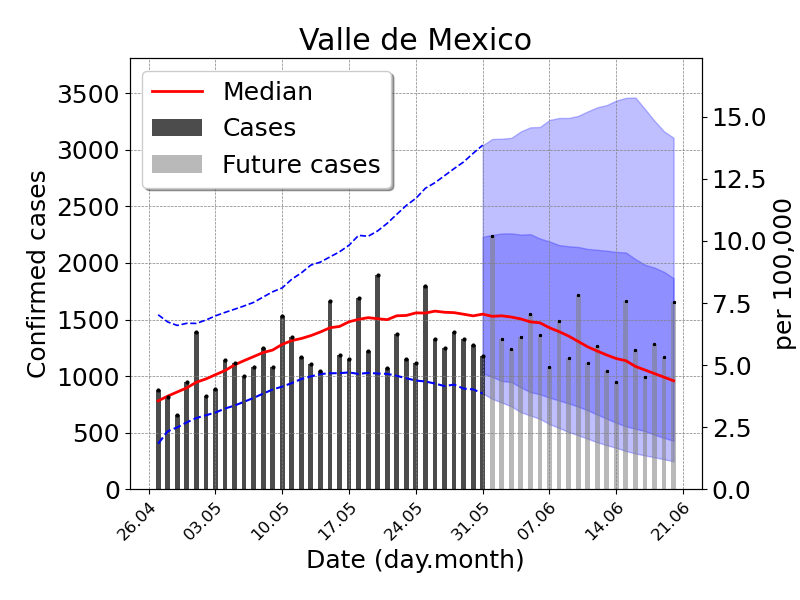} 
 \includegraphics[scale=0.35]{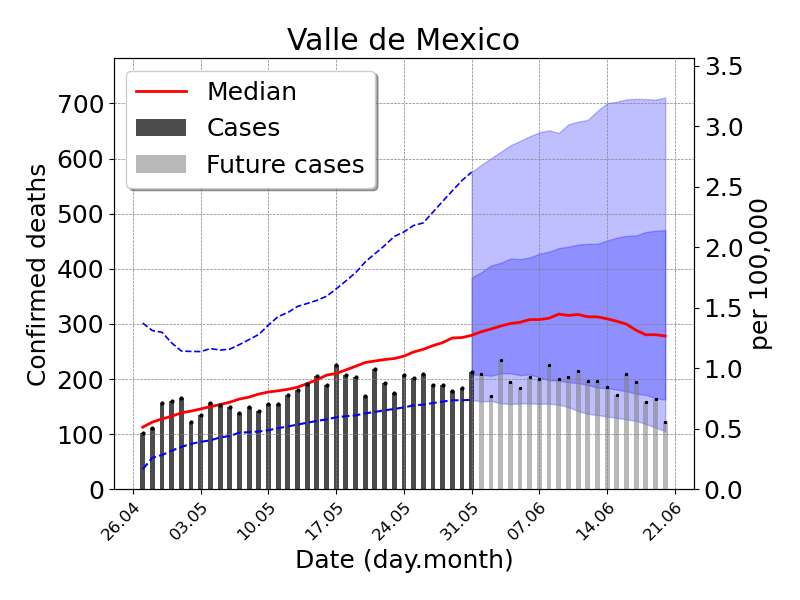}}
 \subfigure[The previous forecastings were superimposed.]{
  \includegraphics[scale=0.35]{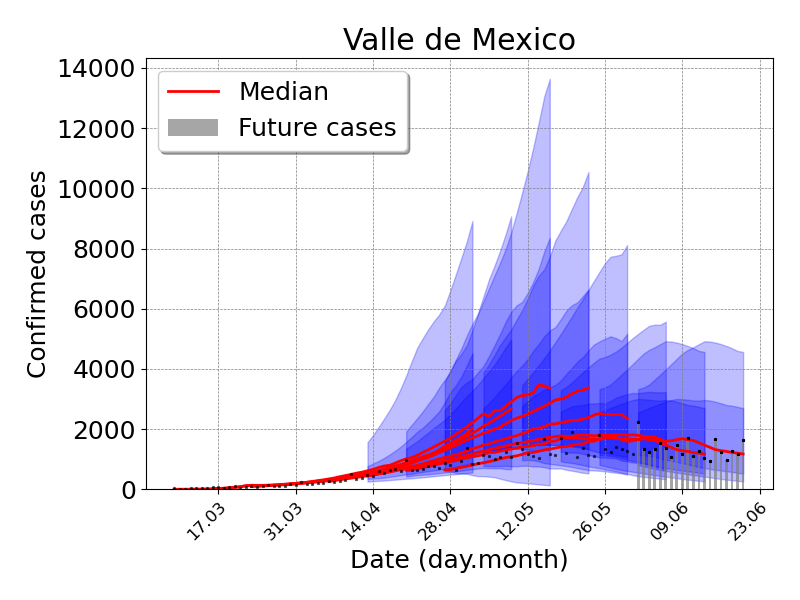} 
 \includegraphics[scale=0.35]{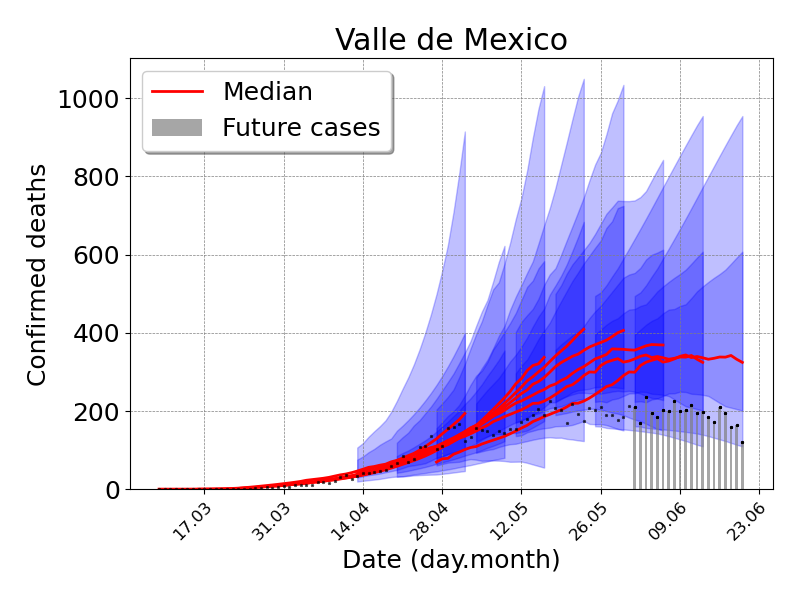}}
 
\caption{Outbreak analysis for Mexico city metropolitan area. From left to right, confirmed cases and deaths. 
\label{fig:forecast7_VM}}
\end{figure}

\begin{figure}[H]
\subfigure[Forecast 14, using data from June 15 to July 19]{
 \includegraphics[scale=0.35]{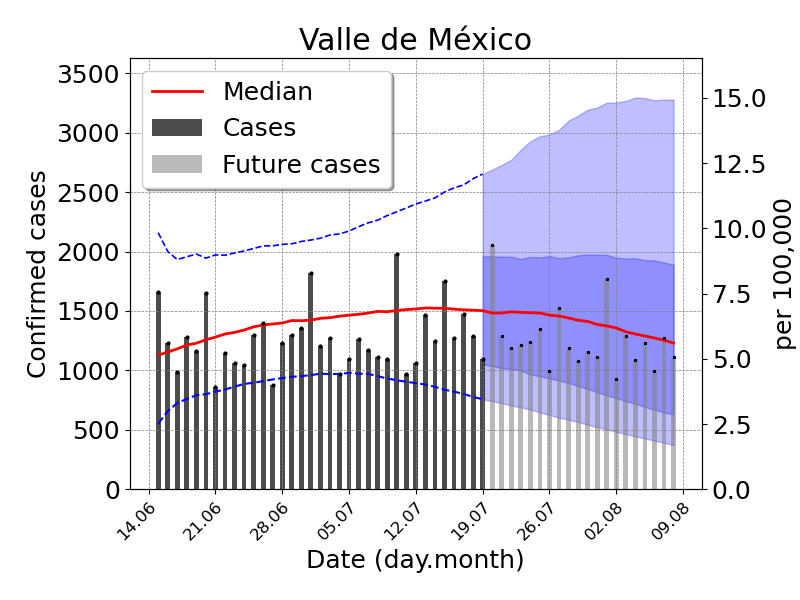} 
 \includegraphics[scale=0.35]{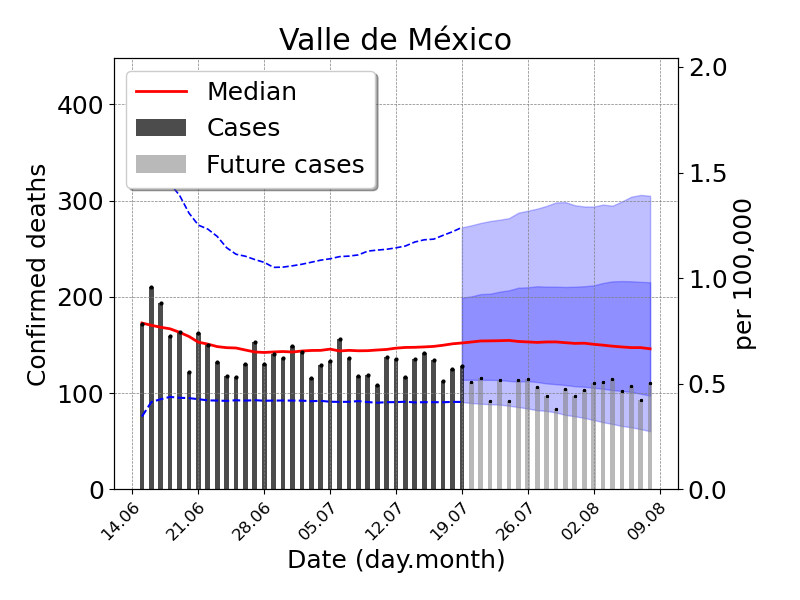}}
 \subfigure[The previous forecastings were superimposed.]{
  \includegraphics[scale=0.35]{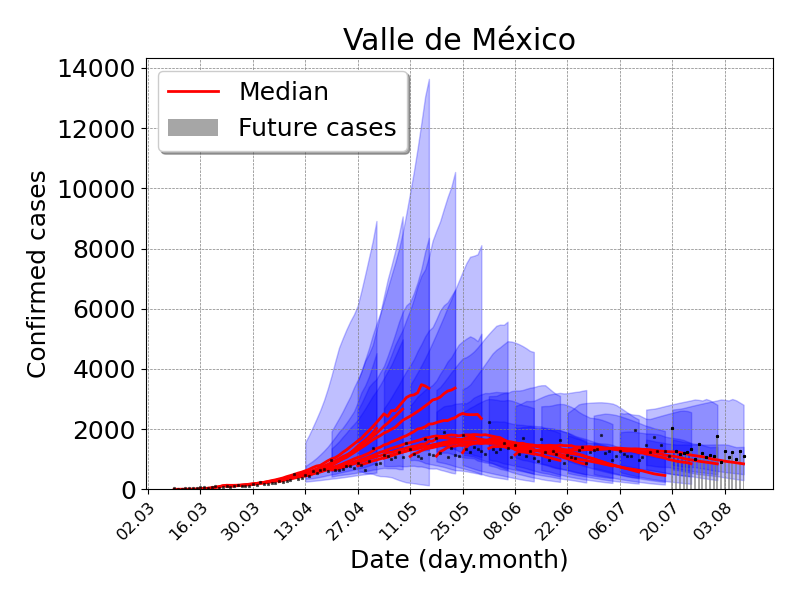} 
 \includegraphics[scale=0.35]{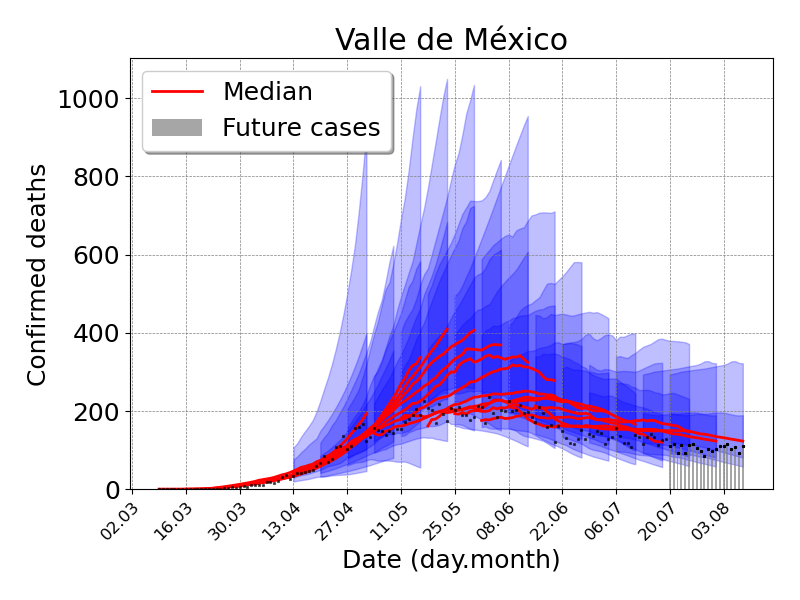}}
 
\caption{Outbreak analysis for Mexico city metropolitan area. From left to right, confirmed cases and deaths. Central red lines indicate the median incidence forecast. The darker shaded region indicates the interquartile forecast range, and the lighter shaded region indicates the 5–95th percentile range. All displayed forecast durations are 20 days from the point of prediction. \label{fig:forecast14_VM}}
\end{figure}

\begin{figure}[H]
\subfigure[Forecast 40, using data from December 14 to January 17]{
 \includegraphics[scale=0.35]{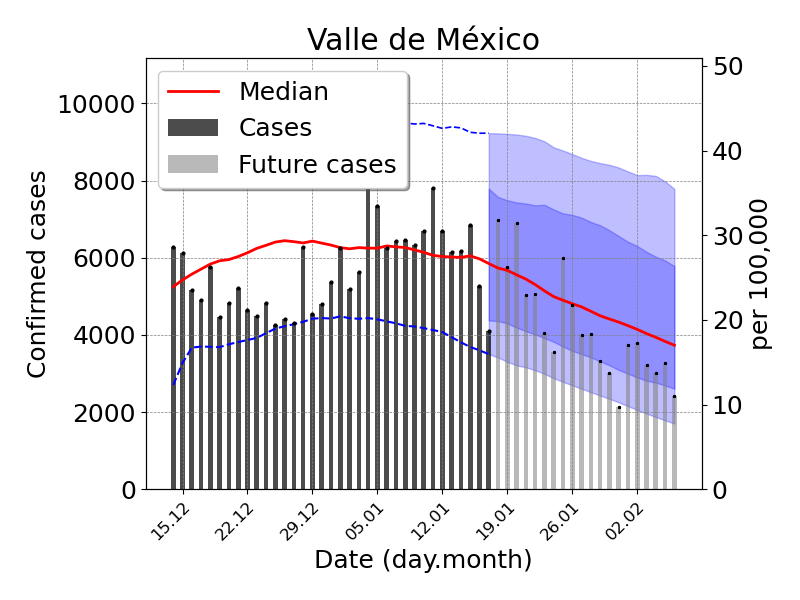} 
 \includegraphics[scale=0.35]{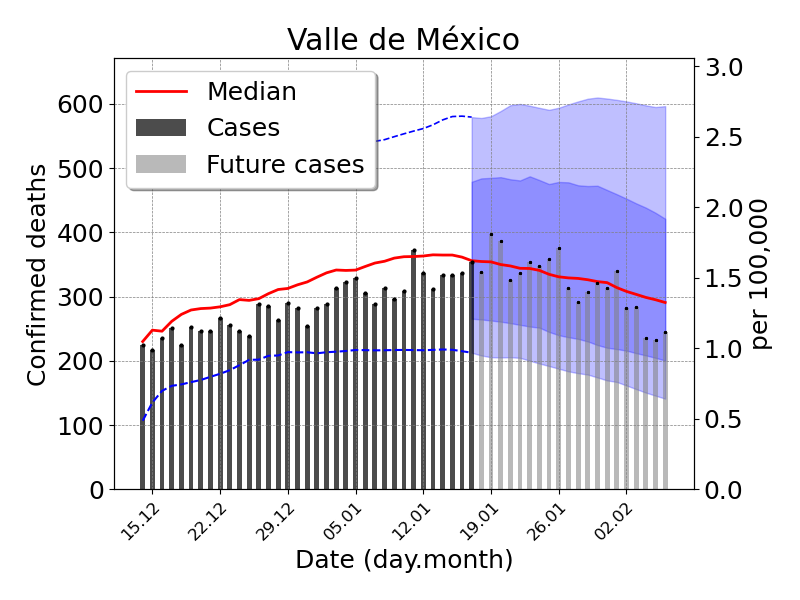}}
\subfigure[The previous forecastings were superimposed.]{ 
  \includegraphics[scale=0.35]{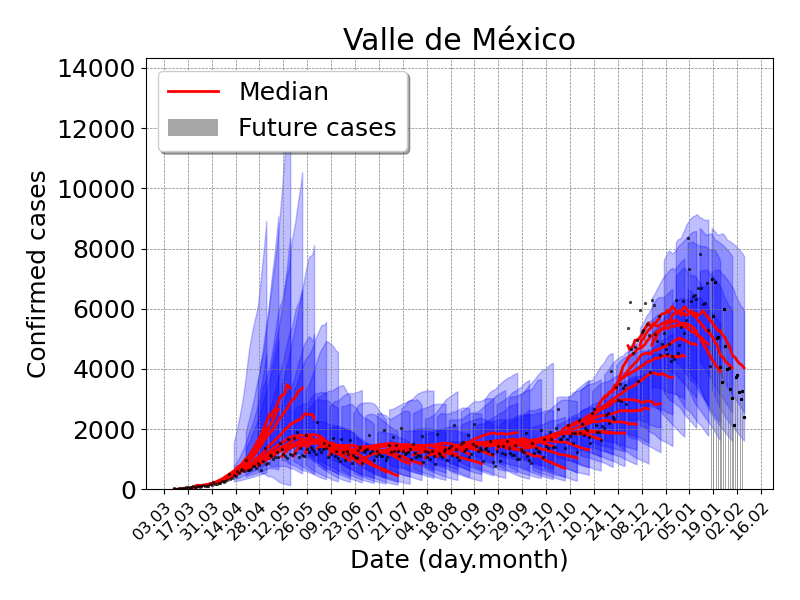} 
 \includegraphics[scale=0.35]{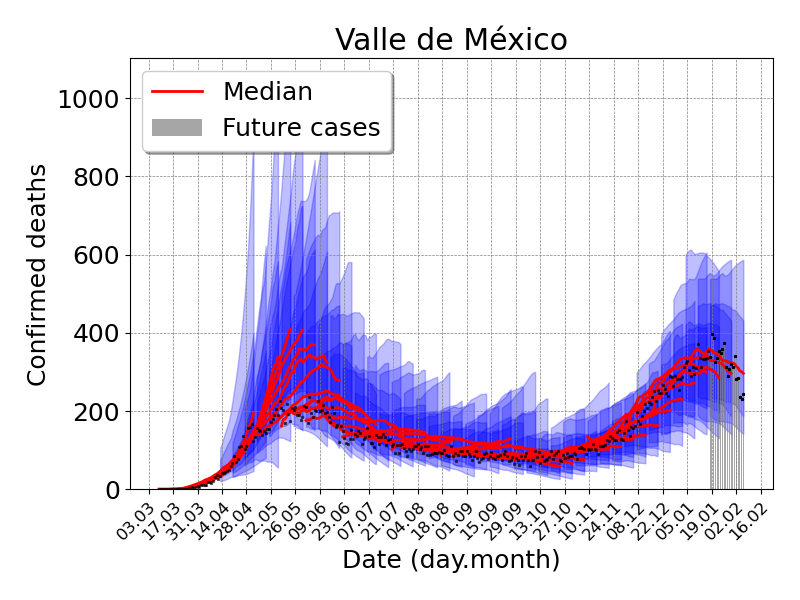}}
\caption{Outbreak analysis for Mexico city metropolitan area. From left to right, confirmed cases and deaths. Central red lines indicate the median incidence forecast. The darker shaded region indicates the interquartile forecast range, and the lighter shaded region indicates the 5–95th percentile range. All displayed forecast durations are 20 days from the point of prediction. \label{fig:forecast40_VM}}
\end{figure}

\begin{figure}[h!]
 \includegraphics[scale=0.26]{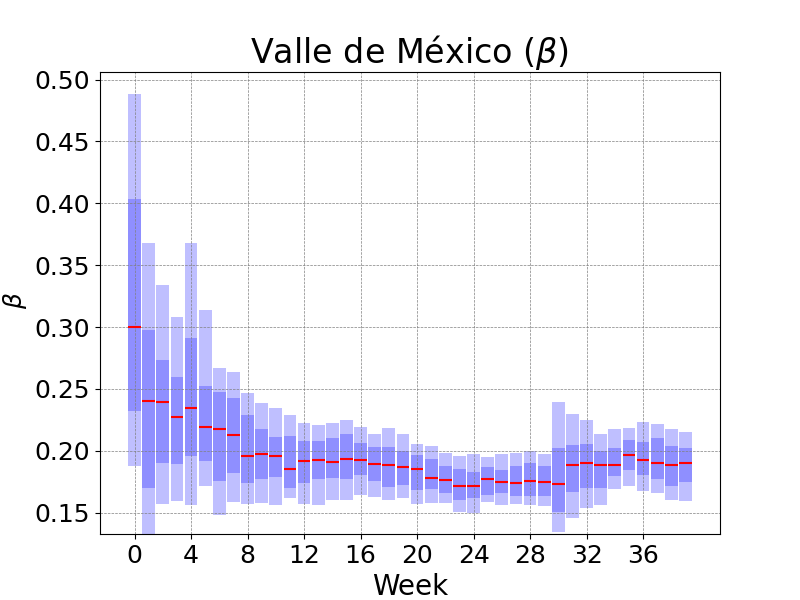}
 \includegraphics[scale=0.26]{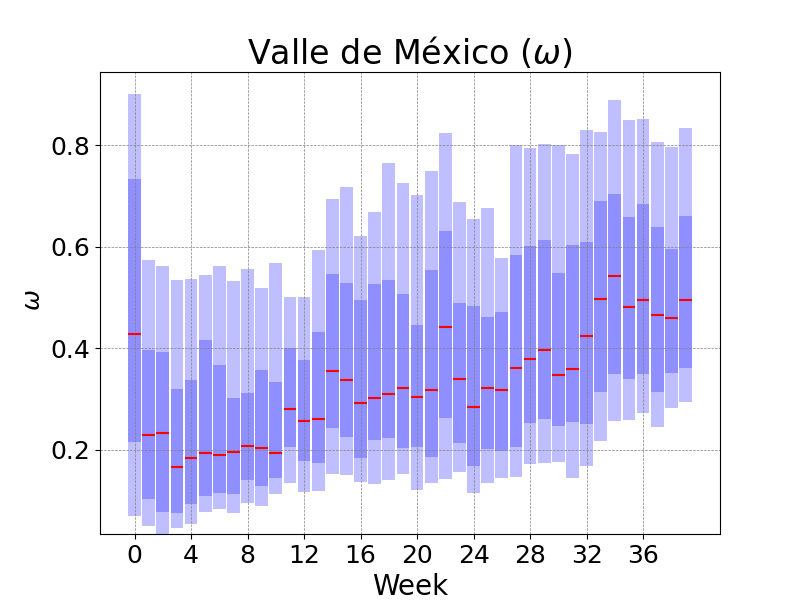}
 \includegraphics[scale=0.26]{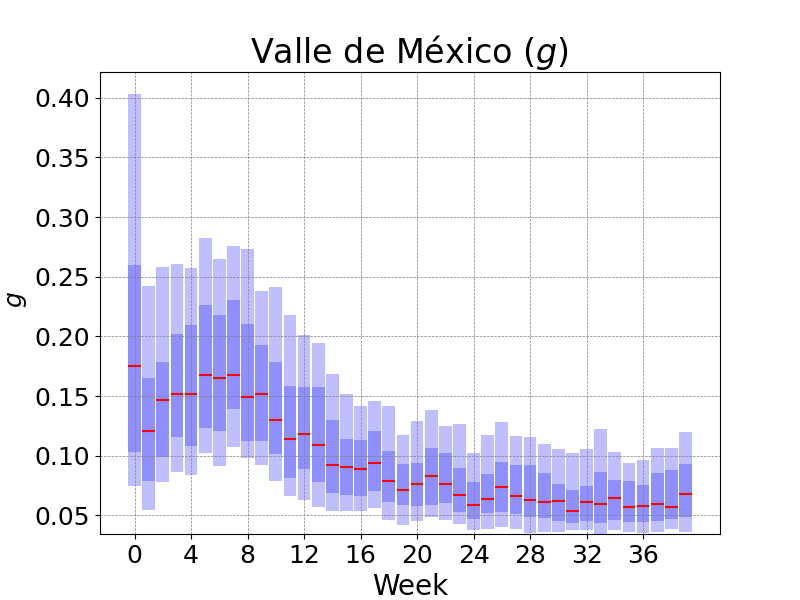}
\caption{Outbreak analysis for Mexico city metropolitan area. From left to right, contact rate after lockdown ($\beta$),  proportion of the effective population ($\omega$), and the fraction of infected dying ($g$). Central red lines indicate median incidence forecast. Darker shaded region indicates forecast interquartile range, and lighter shaded region indicates 5–95th percentile range. \label{fig:evolution_VM}}
\end{figure}

The next predictions using  previous information and update data quickly increase accuracy, i.e., the median of the forecast becomes close to the future data, and the cone of uncertainty shrinks, Figures \ref{fig:forecast7_VM}-\ref{fig:forecast14_VM}. These results point to opportunities in capturing behavioral changes during sequential forecasts and predicting in real-time the future trajectory of the disease after sufficient observations. Figures \ref{fig:forecast7_VM}-\ref{fig:forecast40_VM} show how the model captures the tendency changes in the pandemic dynamic. Also, changes in the parameters associated with the public health measures such as the contact rates ($\beta$), the proportion of effective population size during the outbreak $(\omega)$, and the proportion of observed infectious dying (g) are exhibited in Figure \ref{fig:evolution_VM}.

\section*{Acknowledgments}
The authors are partially founded by CONACYT CB-2016-01-284451 grant. AC was partially supported by UNAM PAPPIT–IN106118 grant. MLDT was funded by FORDECYT 296737 “CONSORCIO EN INTELIGENCIA ARTIFICIAL".

\newpage
%\section{appendix}
\bibliographystyle{unsrt}
\bibliography{final_size.bib} 
\newpage

\subsection{Other examples}
\subsubsection*{Guanajuato}
\begin{figure}[H]
\subfigure[Forecast 7, using data from May 11 to June 14]{
 \includegraphics[scale=0.35]{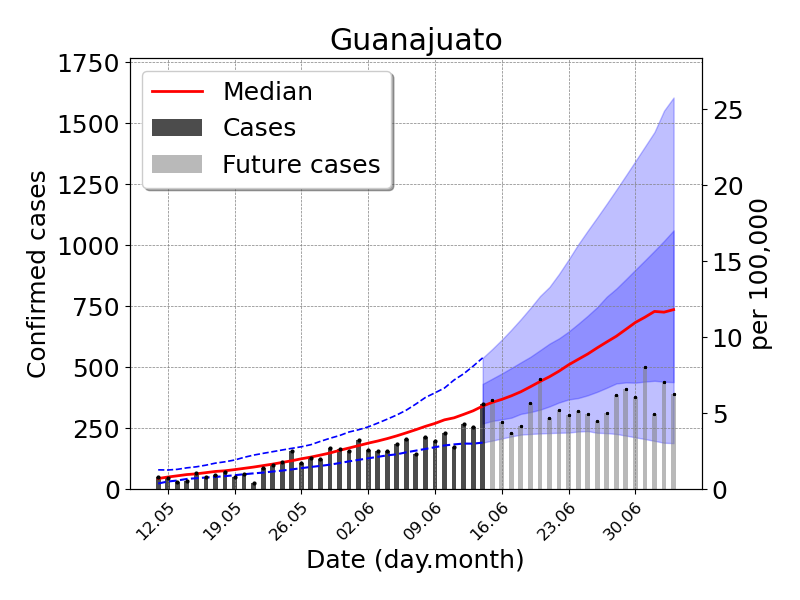} 
 \includegraphics[scale=0.35]{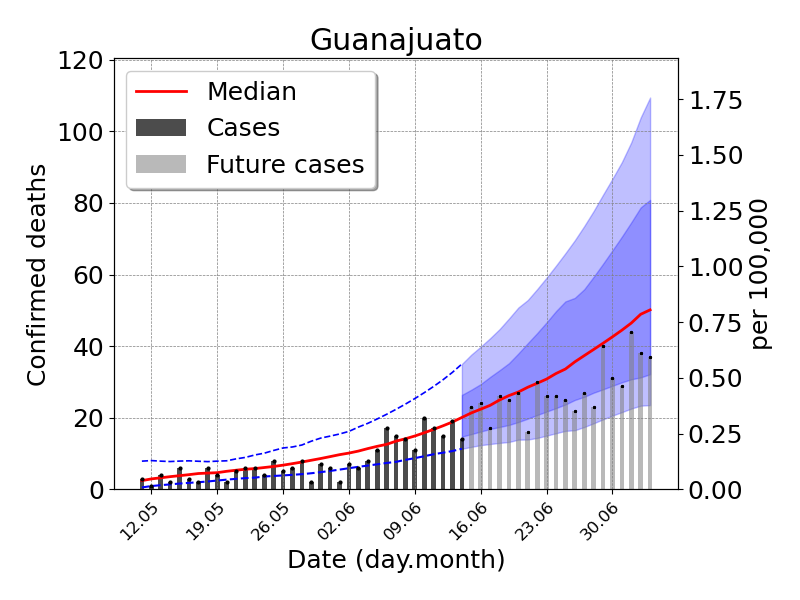}}
 \subfigure[The previous forecastings were superimposed..]{
  \includegraphics[scale=0.35]{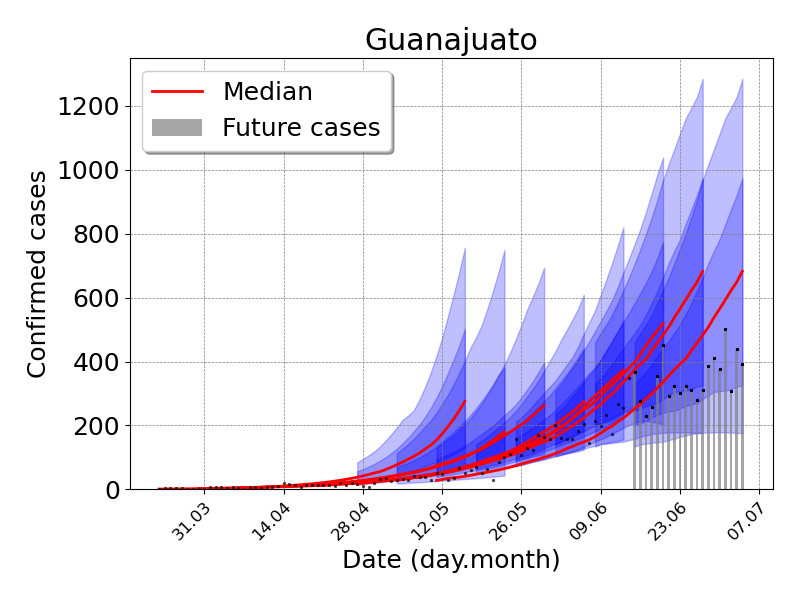} 
 \includegraphics[scale=0.35]{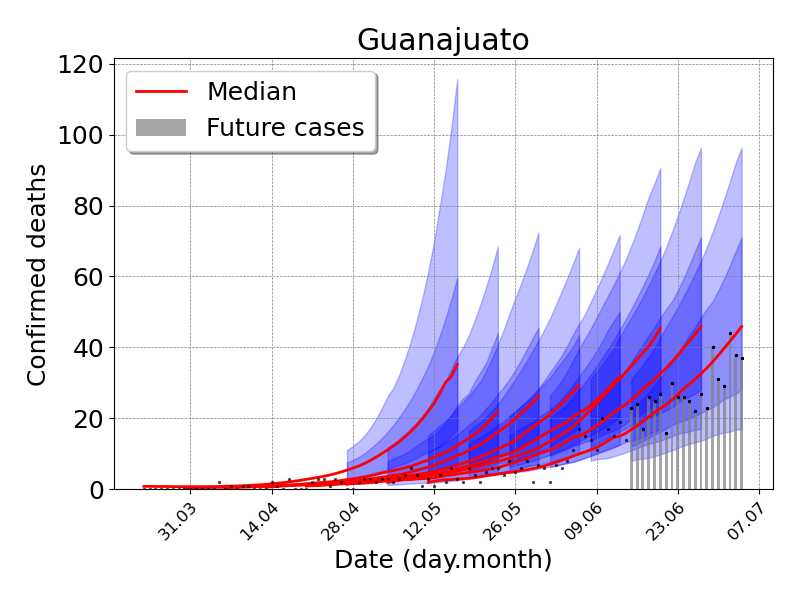}}
 
\caption{Outbreak analysis for Guanajuato. From left to right, confirmed cases and deaths. Central red lines indicate the median incidence forecast. The darker shaded region indicates the interquartile forecast range, and the lighter shaded region indicates the 5–95th percentile range. All displayed forecast durations are 20 days from the point of prediction. Total population 6,228,175 inhabitants.}
\end{figure}

\begin{figure}[H]
\subfigure[Forecast 14, using data from June 27 to August 2.]{
 \includegraphics[scale=0.35]{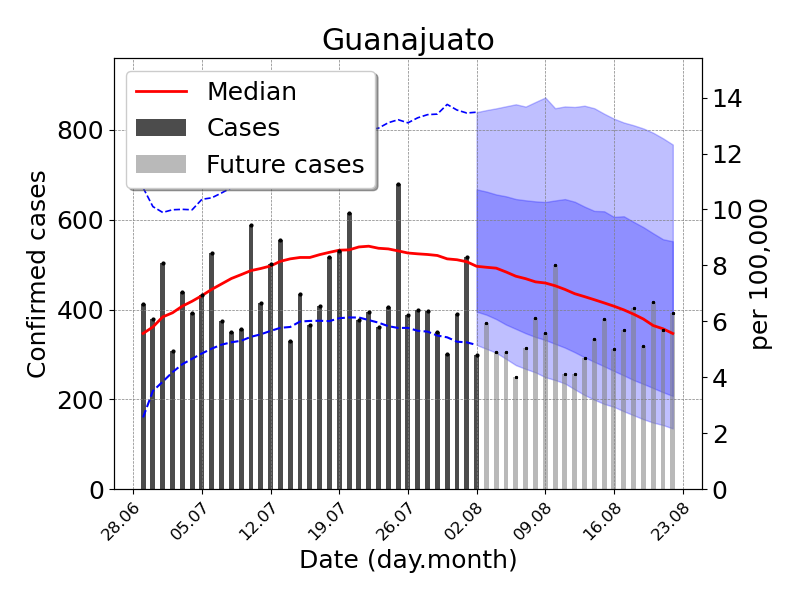} 
 \includegraphics[scale=0.35]{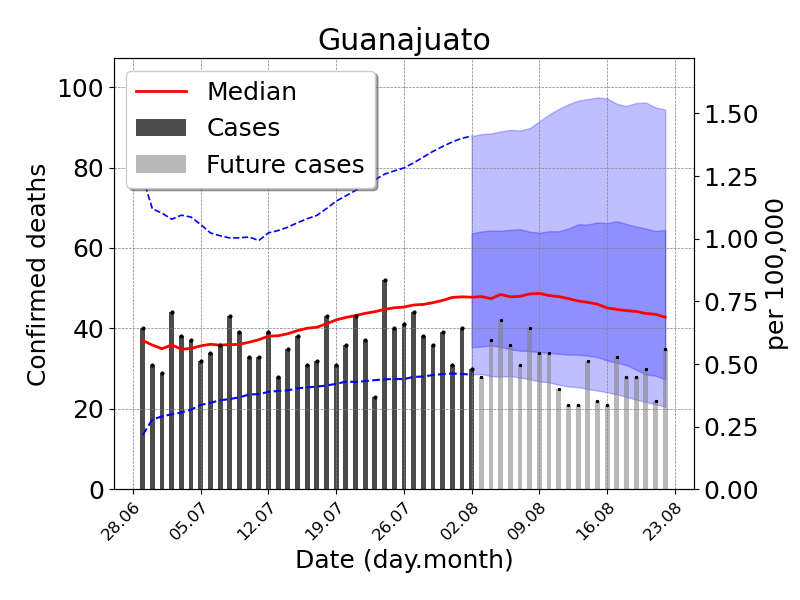}}
 \subfigure[The previous forecastings were superimposed.]{
  \includegraphics[scale=0.35]{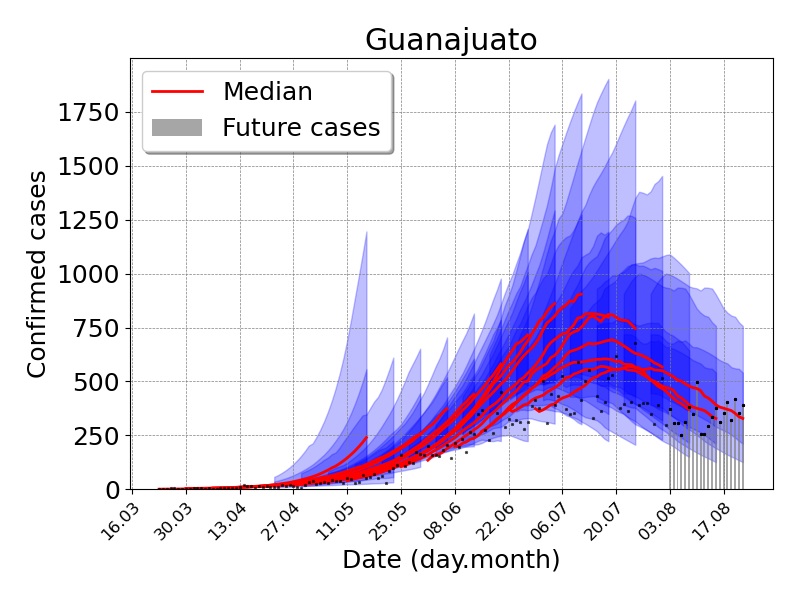} 
 \includegraphics[scale=0.35]{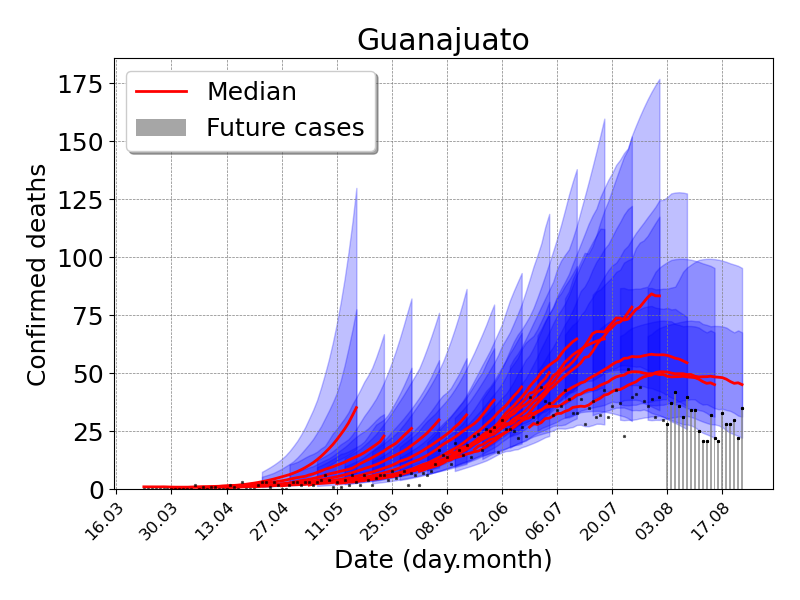}}
 
\caption{Outbreak analysis for Guanajuato. From left to right, confirmed cases and deaths. Central red lines indicate the median incidence forecast. The darker shaded region indicates the interquartile forecast range, and the lighter shaded region indicates the 5–95th percentile range. All displayed forecast durations are 20 days from the point of prediction. }
\end{figure}

\begin{figure}[H]
\subfigure[Forecast 14, using data from December 14 to January 17.]{
 \includegraphics[scale=0.35]{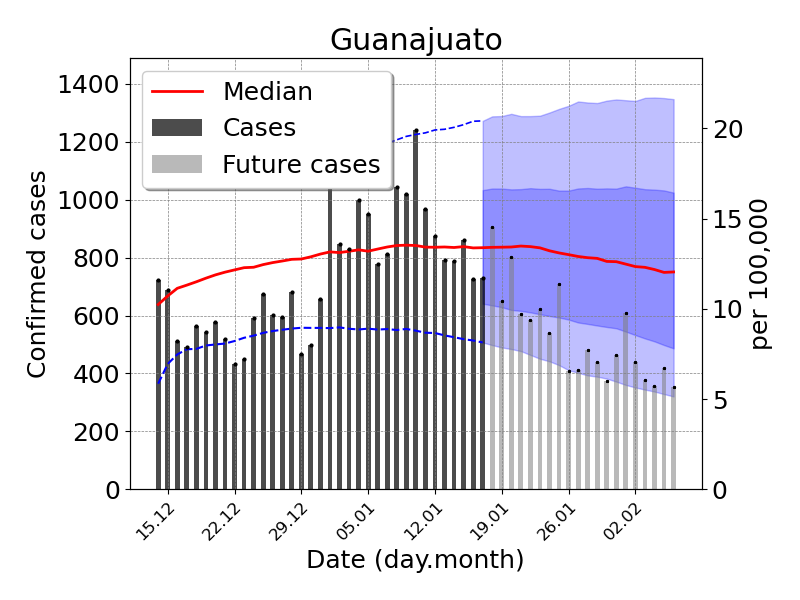} 
 \includegraphics[scale=0.35]{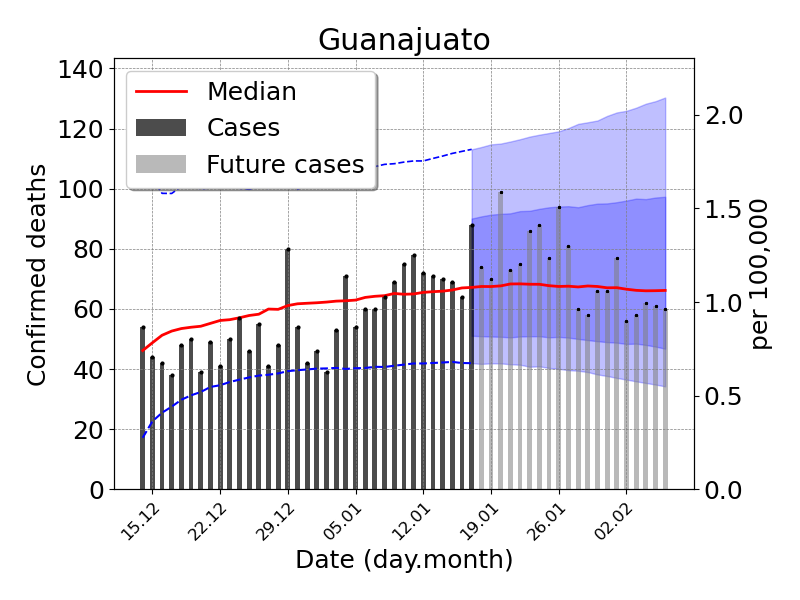}}
 \subfigure[The previous forecastings were superimposed.]{
  \includegraphics[scale=0.35]{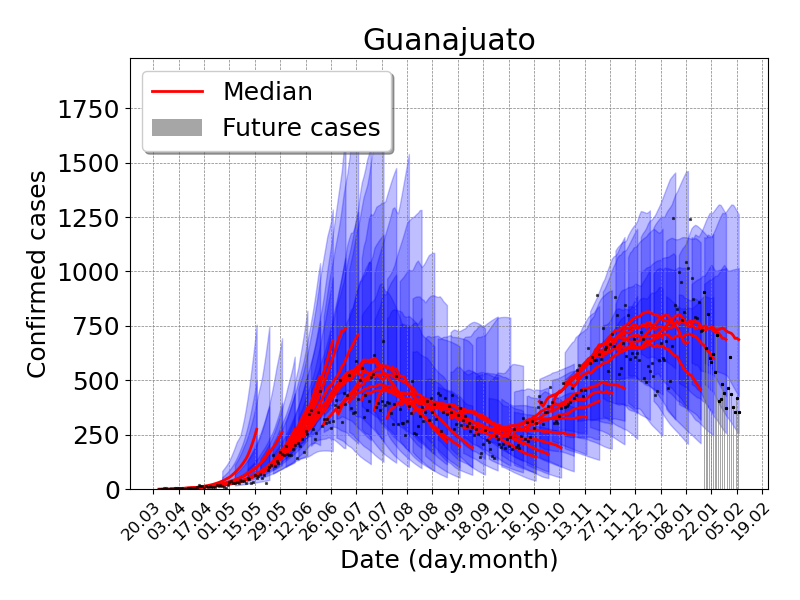} 
 \includegraphics[scale=0.35]{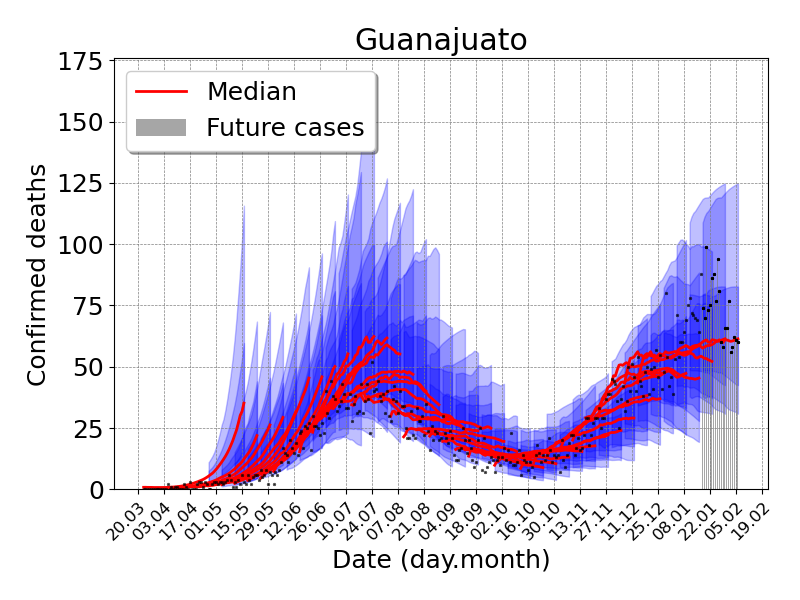}}
 
\caption{Outbreak analysis for Guanajuato. From left to right, confirmed cases and deaths. Central red lines indicate the median incidence forecast. The darker shaded region indicates the interquartile forecast range, and the lighter shaded region indicates the 5–95th percentile range. All displayed forecast durations are 20 days from the point of prediction.}
\end{figure}

\begin{figure}[H]
 \includegraphics[scale=0.26]{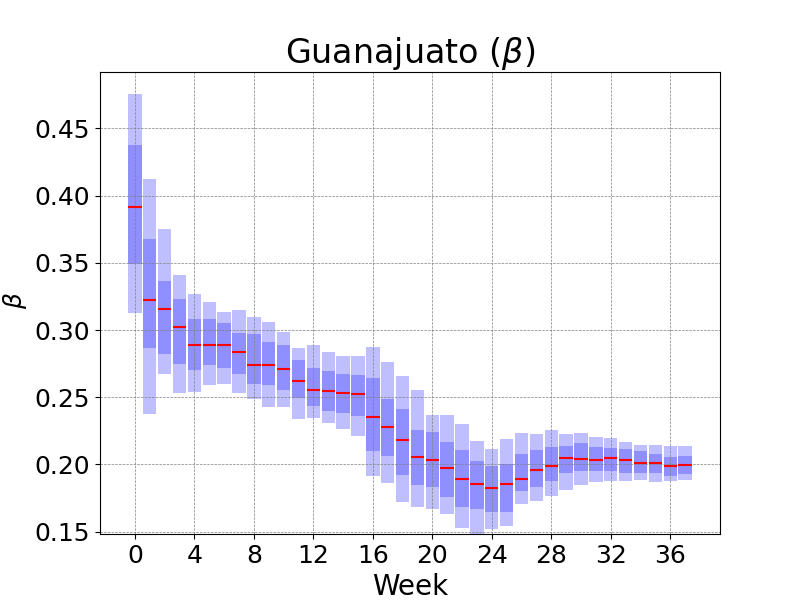}
 \includegraphics[scale=0.26]{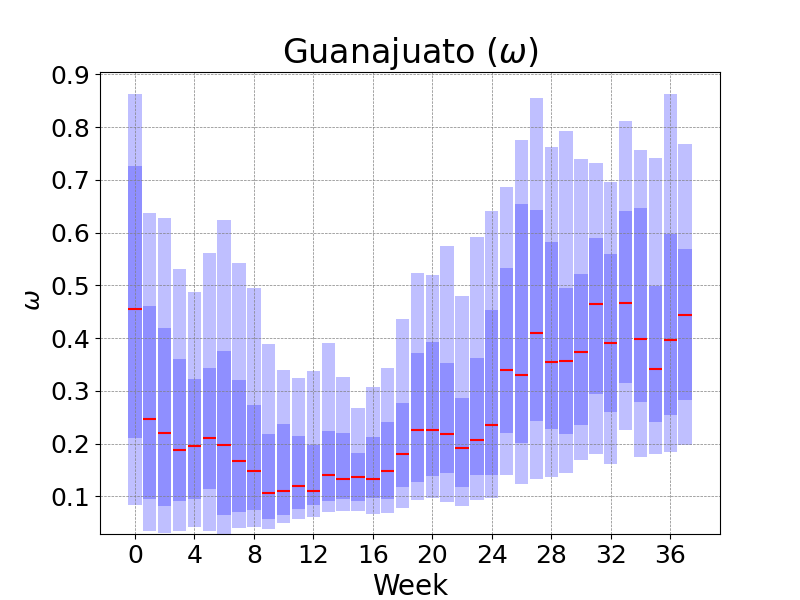}
\includegraphics[scale=0.26]{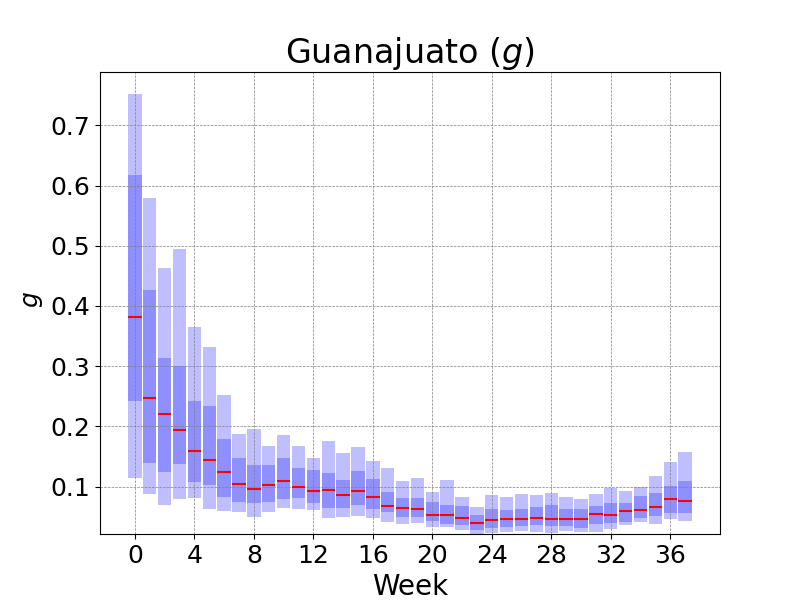}
\caption{Outbreak analysis for Guanajuato. From left to right, contact rate after lockdown ($\beta$),  proportion of the effective population ($\omega$), and the fraction of infected dying ($g$). Central red lines indicate median incidence forecast. Darker shaded region indicates forecast interquartile range, and lighter shaded region indicates 5–95th percentile range.}
\end{figure}

%%%%%%%%%%%%%%%%%%%%%%%%%%%%%%%%%%%%%%%%%%%%%%%%%%%%%%%%%%%%%%%%%%%%%
\subsubsection*{Juaréz}
\begin{figure}[H]
\subfigure[Forecast 7, using data from May 11 to June 14.]{
 \includegraphics[scale=0.35]{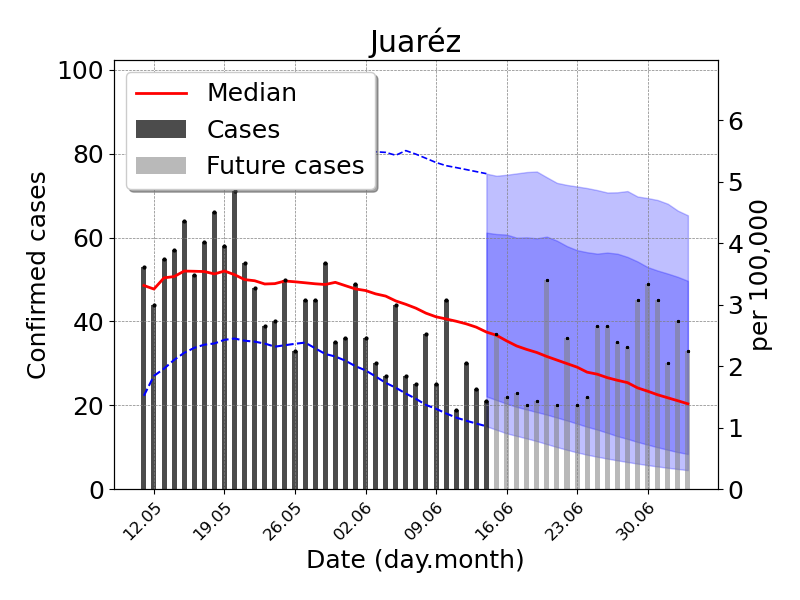} 
 \includegraphics[scale=0.35]{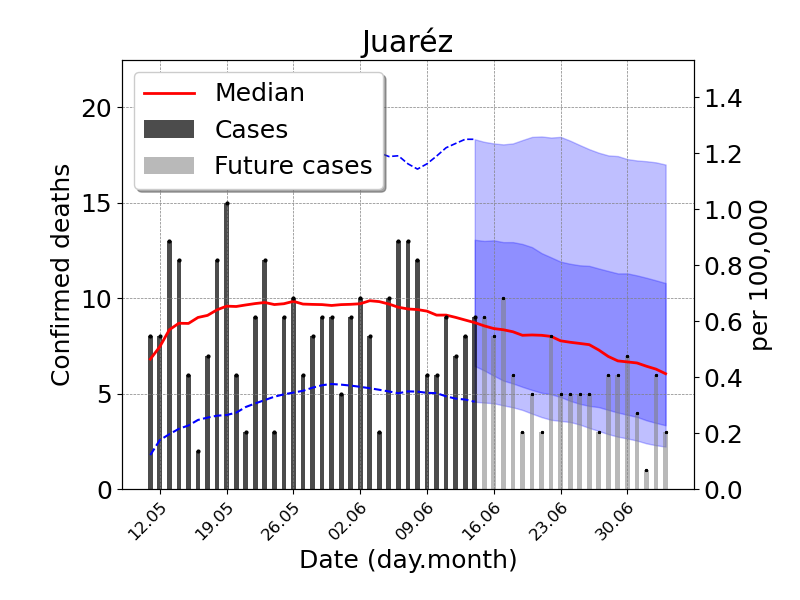}}
 \subfigure[The previous forecastings were superimposed.]{
  \includegraphics[scale=0.35]{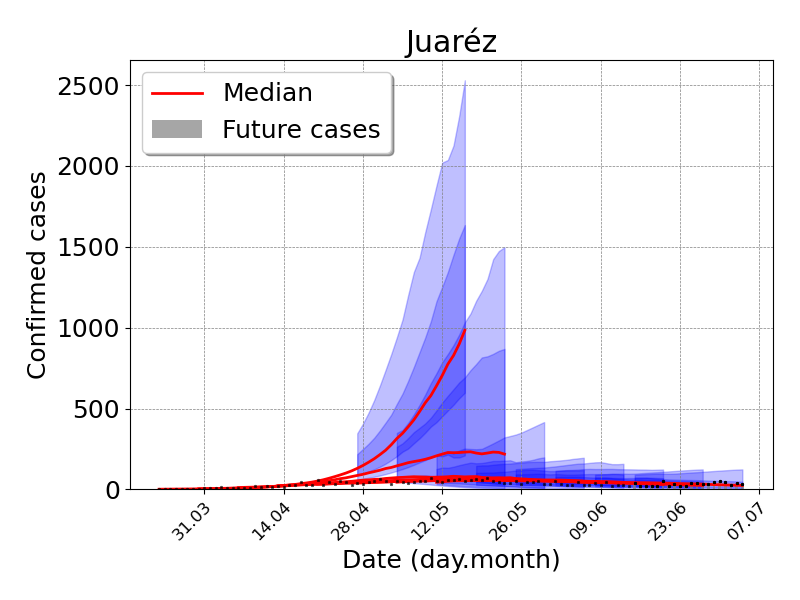} 
 \includegraphics[scale=0.35]{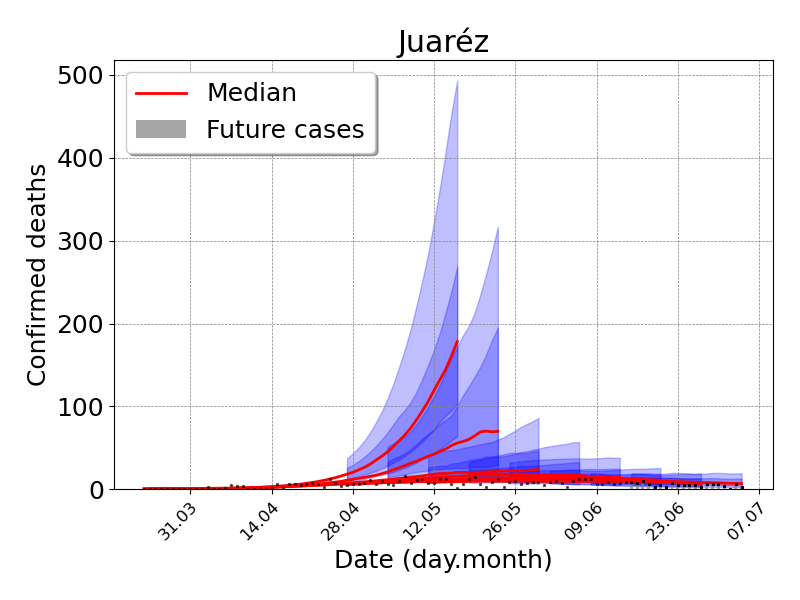}}
 
\caption{Outbreak analysis for Juaréz.  From left to right, confirmed cases and deaths. Central red lines indicate the median incidence forecast. The darker shaded region indicates the interquartile forecast range, and the lighter shaded region indicates the 5–95th percentile range. All displayed forecast durations are 20 days from the point of prediction. Total population 1,464,930 inhabitants.}
\end{figure}

\begin{figure}[H]
\subfigure[Forecast 14, using data from June 29 to August 2]{
 \includegraphics[scale=0.35]{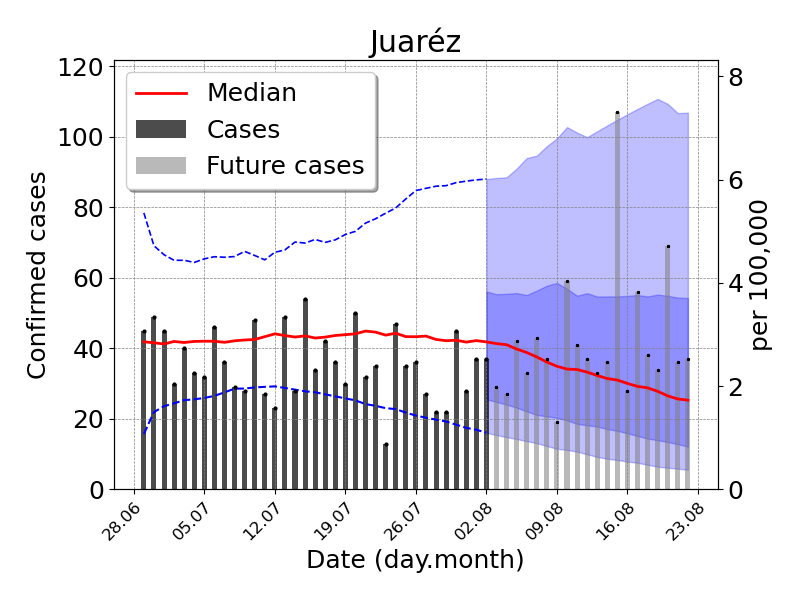} 
 \includegraphics[scale=0.35]{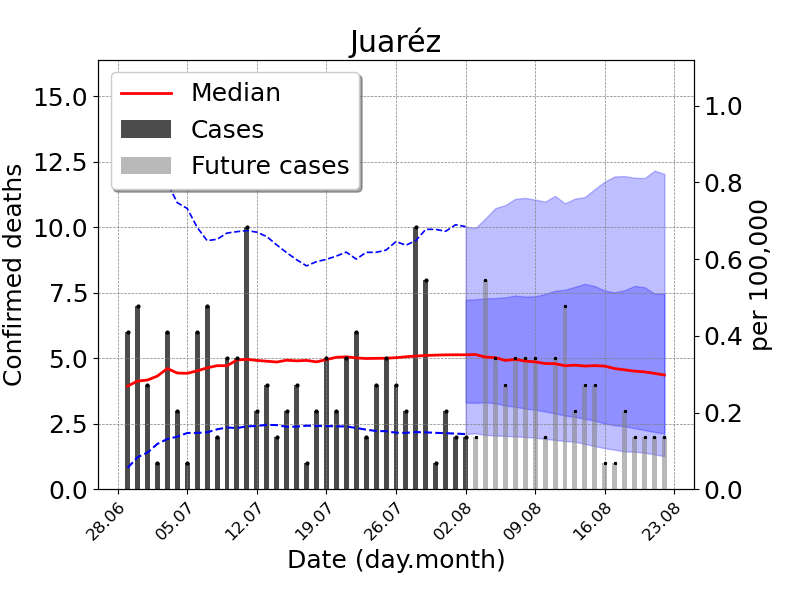}}
 \subfigure[The previous forecastings were superimposed.]{
  \includegraphics[scale=0.35]{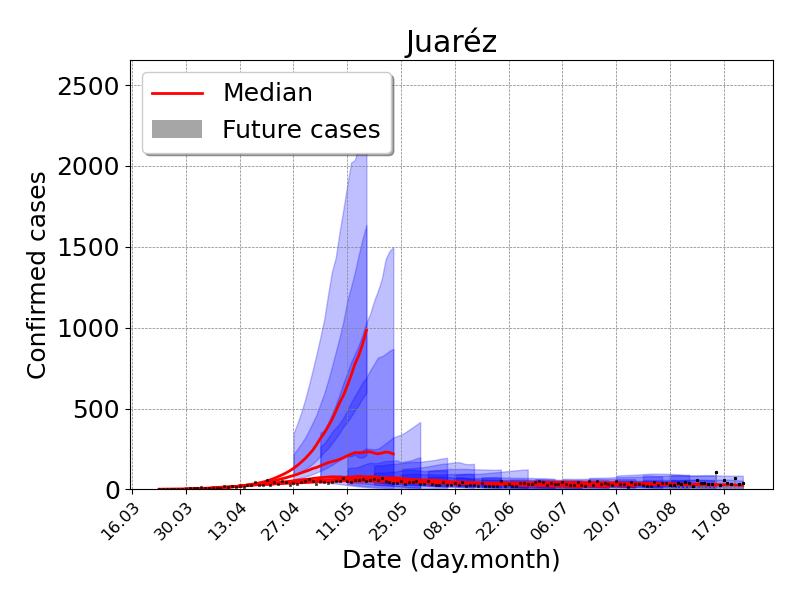} 
 \includegraphics[scale=0.35]{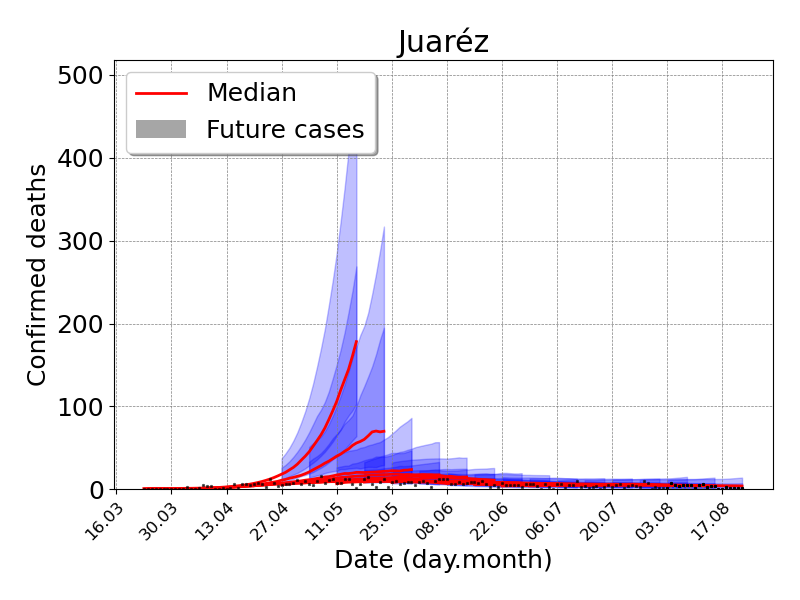}}
 
\caption{Outbreak analysis for Juaréz.  From left to right, confirmed cases and deaths. Central red lines indicate the median incidence forecast. The darker shaded region indicates the interquartile forecast range, and the lighter shaded region indicates the 5–95th percentile range. All displayed forecast durations are 20 days from the point of prediction.}
\end{figure}

\begin{figure}[H]
\subfigure[Forecast 38, using data from December 15 to January 18]{
 \includegraphics[scale=0.35]{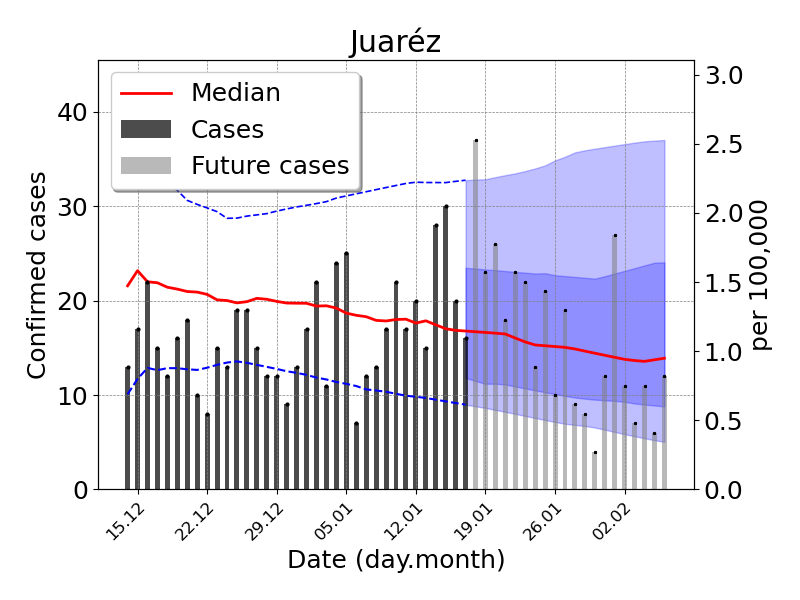} 
 \includegraphics[scale=0.35]{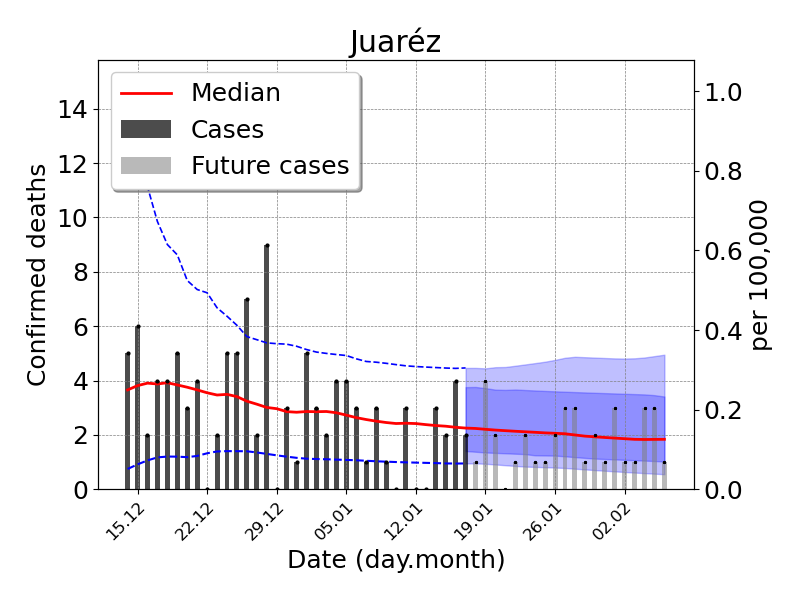}}
 \subfigure[The previous forecastings were superimposed.]{
  \includegraphics[scale=0.35]{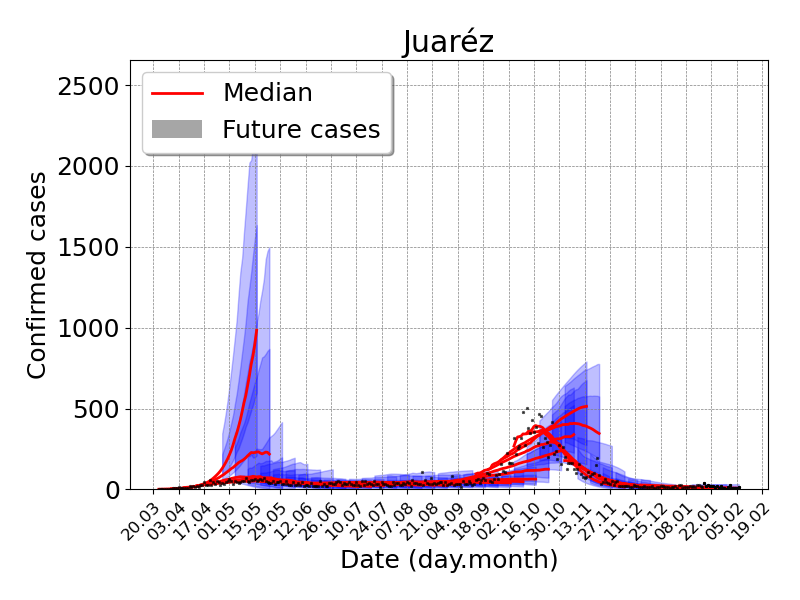} 
 \includegraphics[scale=0.35]{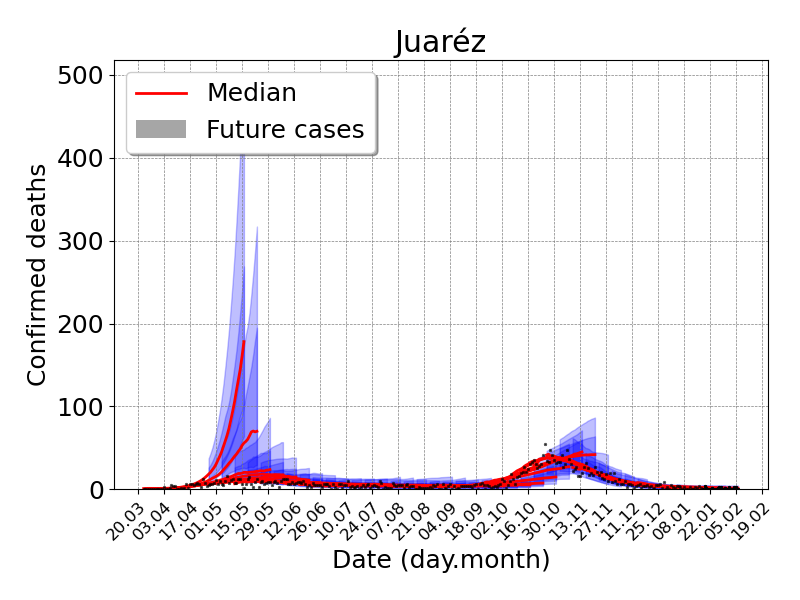}}
 
\caption{Outbreak analysis for Juaréz.  From left to right, confirmed cases and deaths. Central red lines indicate the median incidence forecast. The darker shaded region indicates the interquartile forecast range, and the lighter shaded region indicates the 5–95th percentile range. All displayed forecast durations are 20 days from the point of prediction.}
\end{figure}

\begin{figure}[H]
 \includegraphics[scale=0.26]{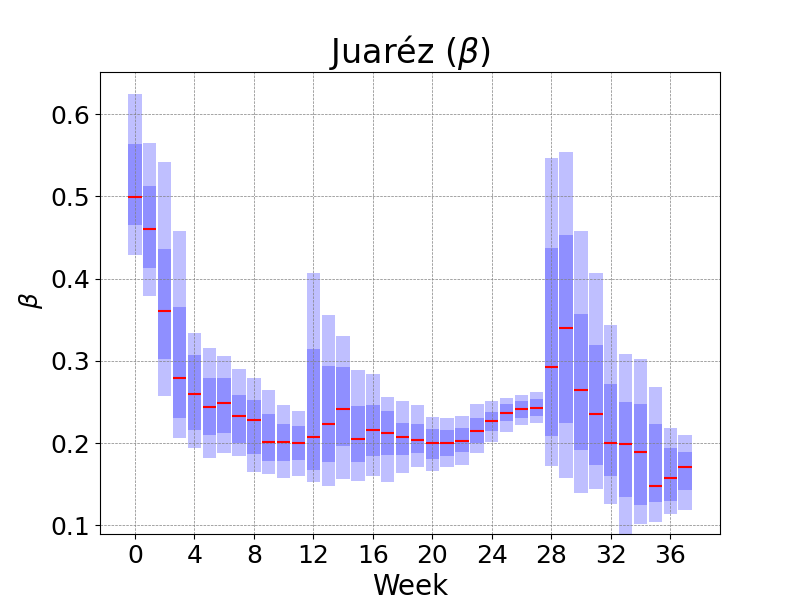}
 \includegraphics[scale=0.26]{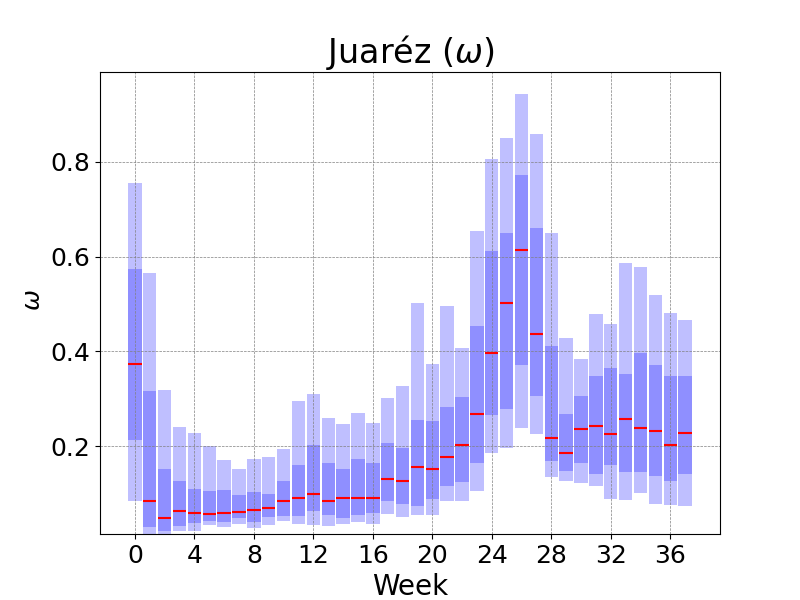}
\includegraphics[scale=0.26]{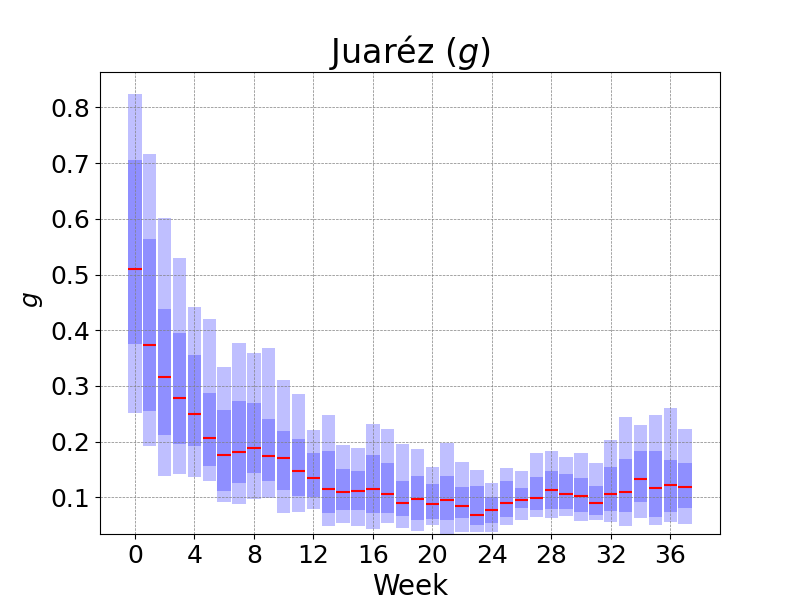}
\caption{Outbreak analysis for Juaréz. From left to right, contact rate after lockdown ($\beta$),  proportion of the effective population ($\omega$), and the fraction of infected dying ($g$). Central red lines indicate median incidence forecast. Darker shaded region indicates forecast interquartile range, and lighter shaded region indicates 5–95th percentile range.}
\end{figure}

%%%%%%%%%%%%%%%%%%%%%%%%%%%%%%%%%%%%%%%%%%%%%%%%%%%%%%%%%%%%%%%%%%%%%
\subsubsection*{Villa Hermosa}
\begin{figure}[H]
\subfigure[Forecast 7, using data from May 11 to June 14.]{
 \includegraphics[scale=0.35]{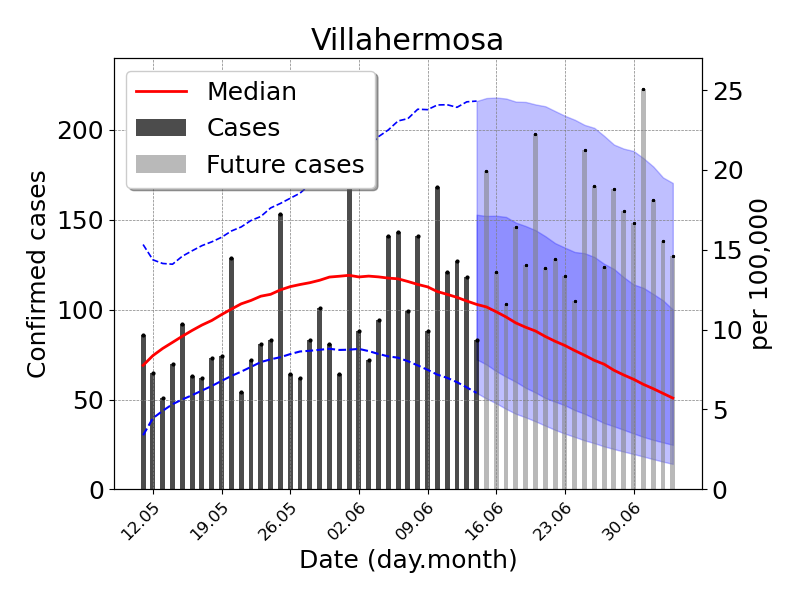} 
 \includegraphics[scale=0.35]{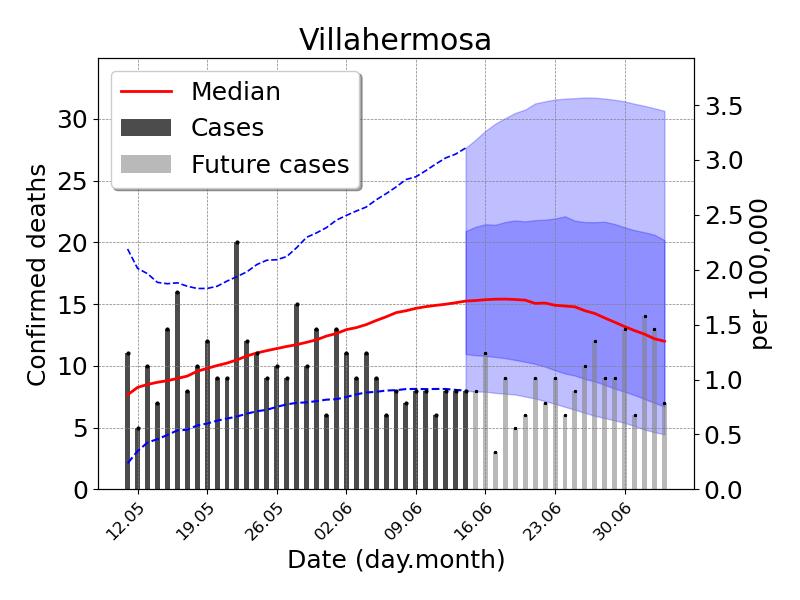}}
 \subfigure[The previous forecastings were superimposed.]{
  \includegraphics[scale=0.35]{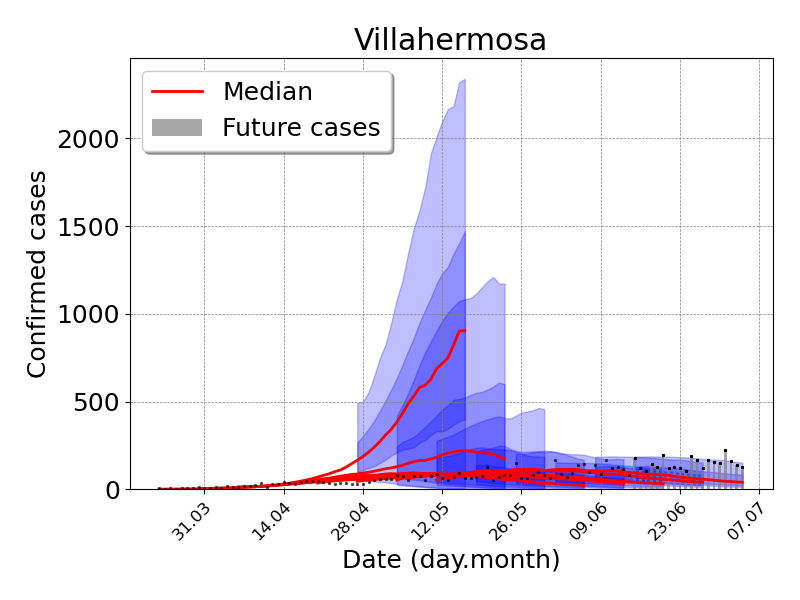} 
 \includegraphics[scale=0.35]{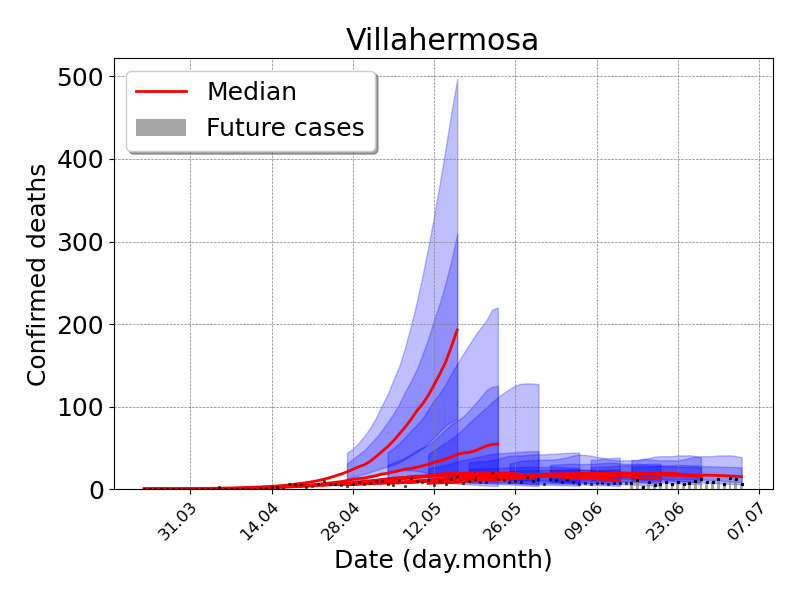}}
 
\caption{Outbreak analysis for Villa Hermosa.  From left to right, confirmed cases and deaths. Central red lines indicate the median incidence forecast. The darker shaded region indicates the interquartile forecast range, and the lighter shaded region indicates the 5–95th percentile range. All displayed forecast durations are 20 days from the point of prediction. Total population 888,867 inhabitants.}
\end{figure}

\begin{figure}[H]
\subfigure[Forecast 14, using data from June 29 to August 2.]{
 \includegraphics[scale=0.35]{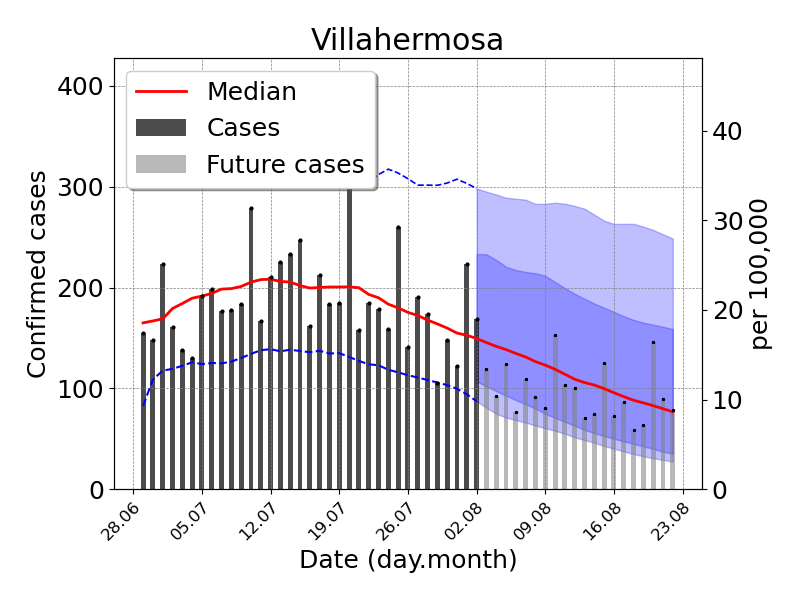} 
 \includegraphics[scale=0.35]{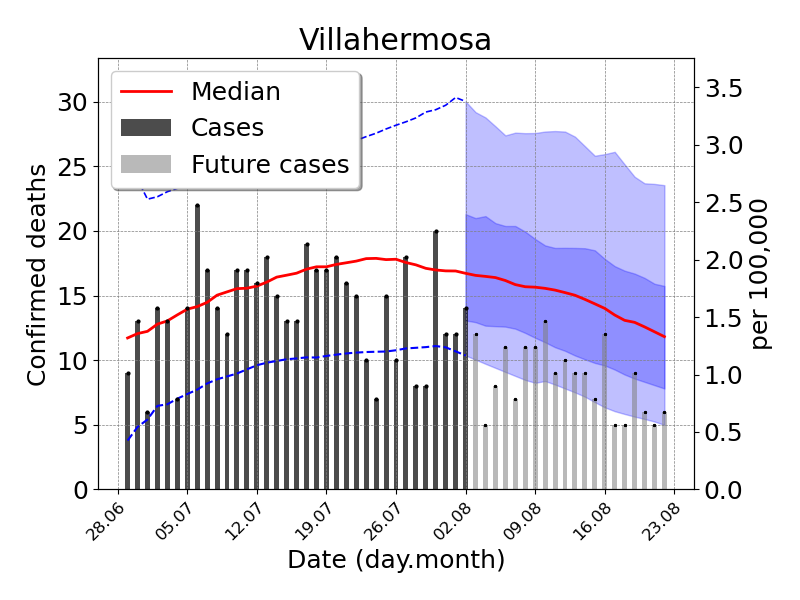}}
 \subfigure[Previous forecast.]{
  \includegraphics[scale=0.35]{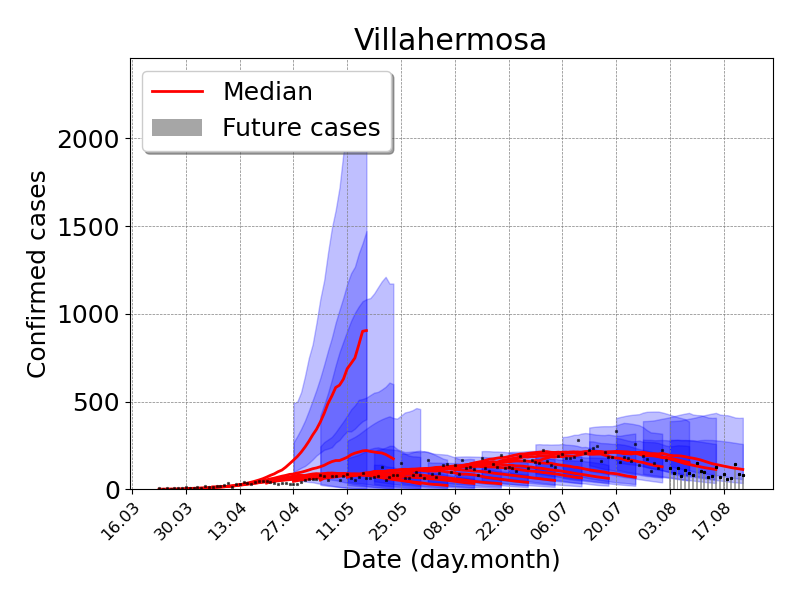} 
 \includegraphics[scale=0.35]{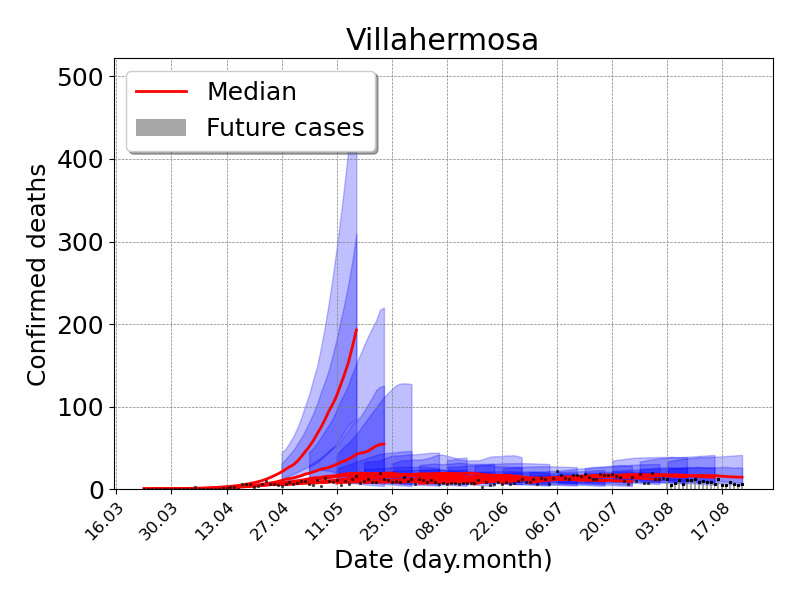}}
 
\caption{Outbreak analysis for Villa Hermosa.  From left to right, confirmed cases and deaths. Central red lines indicate the median incidence forecast. The darker shaded region indicates the interquartile forecast range, and the lighter shaded region indicates the 5–95th percentile range. All displayed forecast durations are 20 days from the point of prediction.}
\end{figure}

\begin{figure}[H]
\subfigure[Forecast 38, using data from December 14 to January 17.]{
 \includegraphics[scale=0.35]{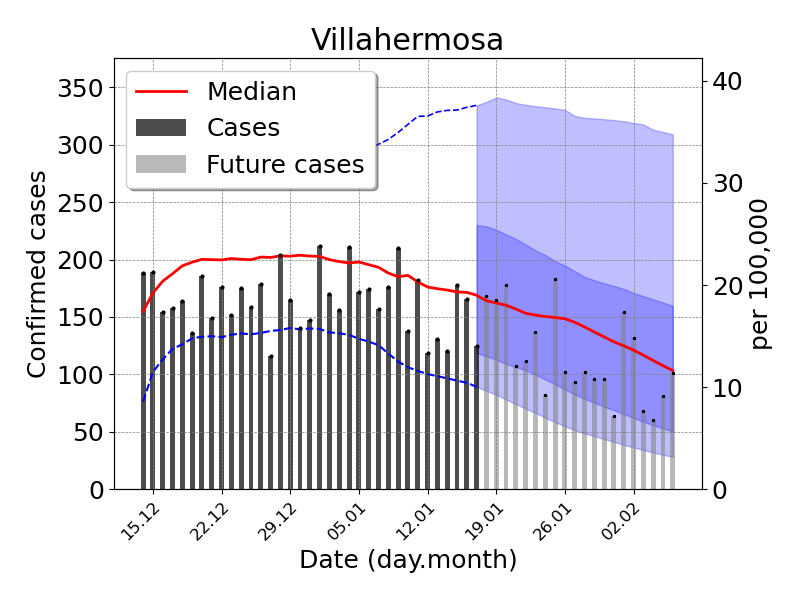} 
 \includegraphics[scale=0.35]{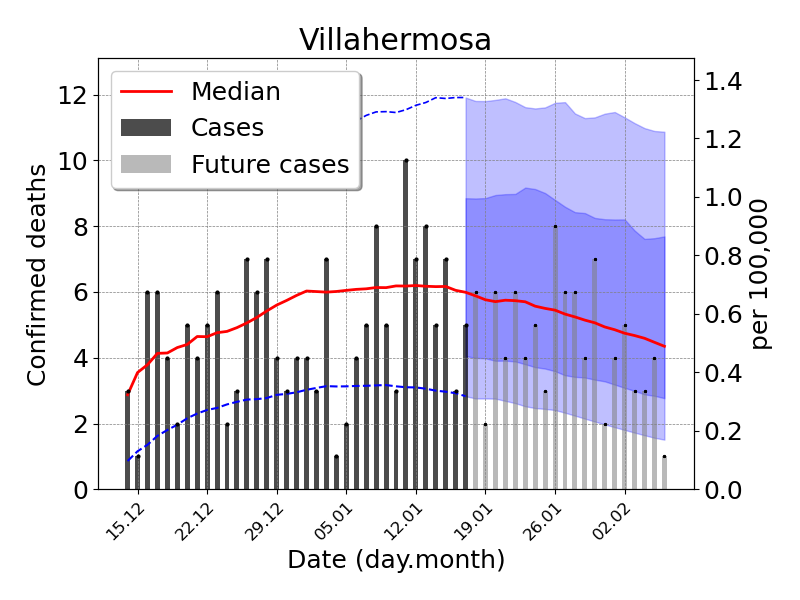}}
 \subfigure[Previous forecast.]{
  \includegraphics[scale=0.35]{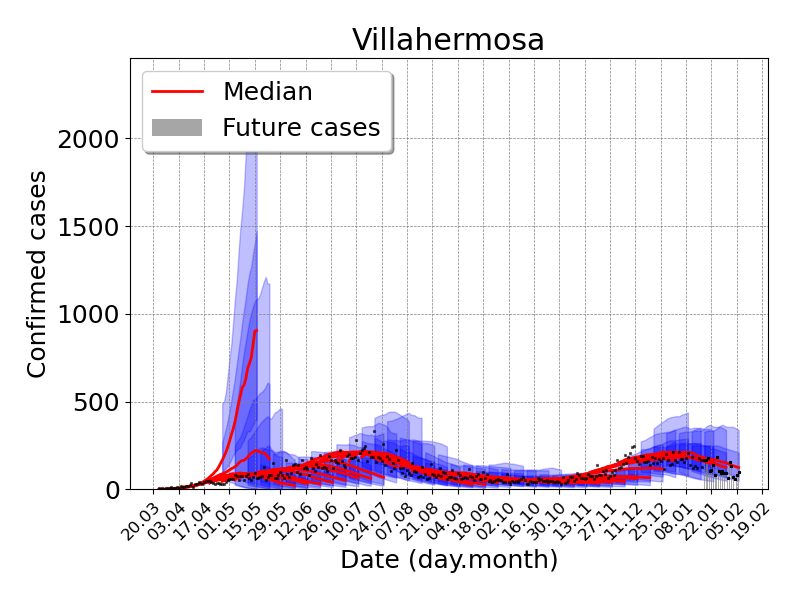} 
 \includegraphics[scale=0.35]{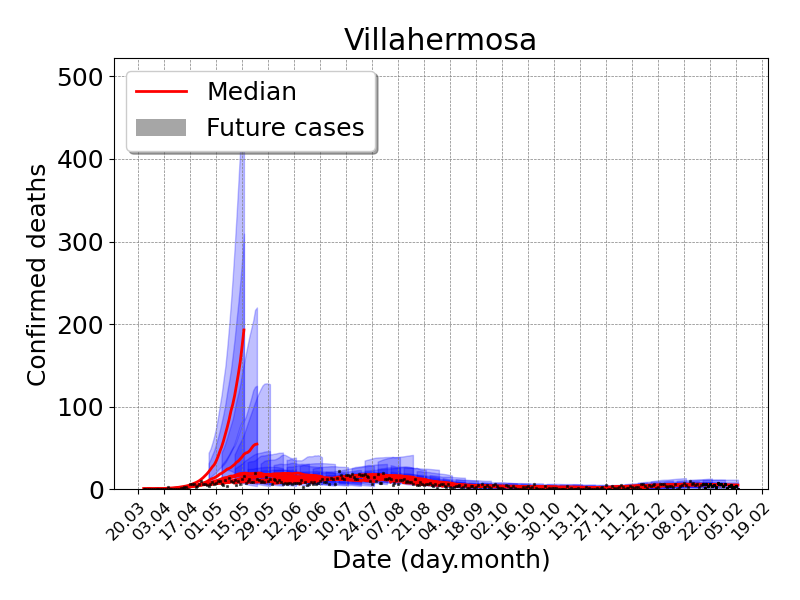}}
 
\caption{Outbreak analysis for Villa Hermosa.  From left to right, confirmed cases and deaths. Central red lines indicate the median incidence forecast. The darker shaded region indicates the interquartile forecast range, and the lighter shaded region indicates the 5–95th percentile range. All displayed forecast durations are 20 days from the point of prediction.}
\end{figure}

\begin{figure}[H]
 \includegraphics[scale=0.26]{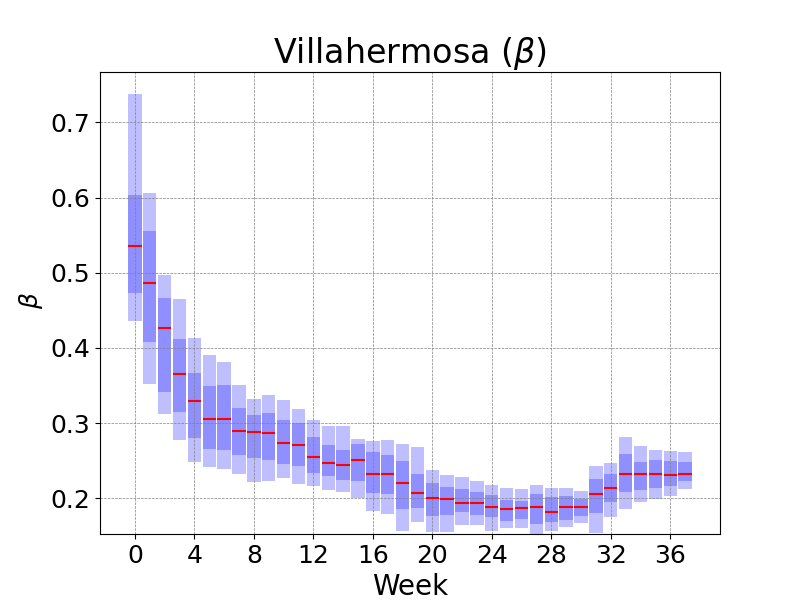}
 \includegraphics[scale=0.26]{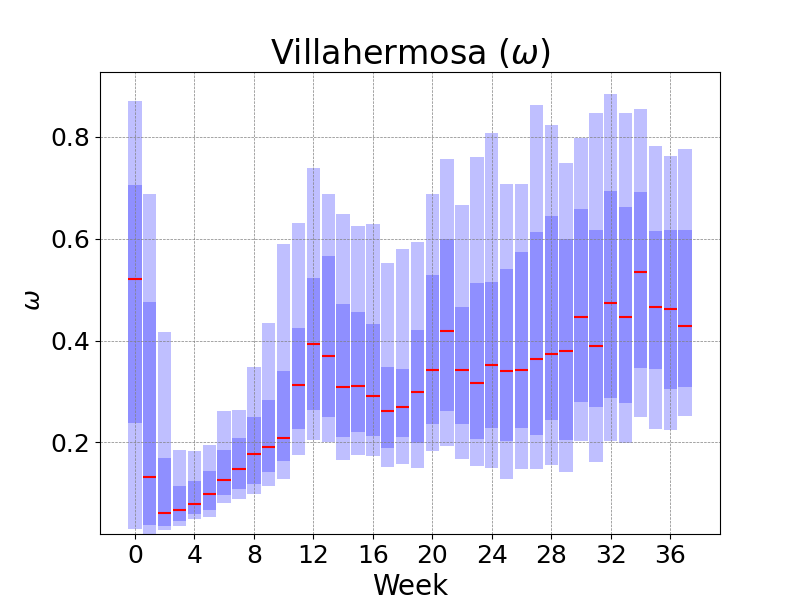}
\includegraphics[scale=0.26]{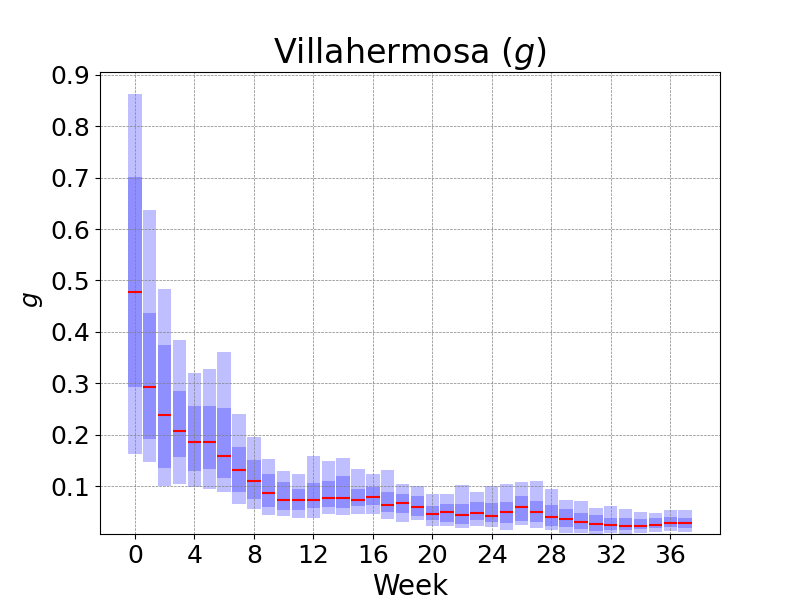}
\caption{Outbreak analysis for Villa Hermosa. From left to right, contact rate after lockdown ($\beta$),  proportion of the effective population ($\omega$), and the fraction of infected dying ($g$). Central red lines indicate median incidence forecast. Darker shaded region indicates forecast interquartile range, and lighter shaded region indicates 5–95th percentile range.}
\end{figure}

\end{document}